\documentclass[aps,pra,showpacs,amssymb,superscriptaddress,twocolumn,nofootinbib]{revtex4-1}
\usepackage[ansinew]{inputenc}
\usepackage{bbm}
\usepackage{bm}
\usepackage{amsbsy}
\usepackage{amsthm}
\usepackage{amssymb}
\usepackage{amsfonts}
\usepackage{amsmath}
\usepackage{dsfont} % for symbols like the identity: \mathds{1}
\usepackage{graphicx} % for graphics
\usepackage{epsfig}
\usepackage{epstopdf}
\usepackage{dsfont}
\usepackage{color}
\usepackage[colorlinks]{hyperref}
\usepackage[figure,table]{hypcap}
\usepackage{enumerate}%allows different styles of enumerate environment
\usepackage{appendix}
\hypersetup{
	bookmarksnumbered,
	pdfstartview={FitH},
	citecolor={darkgreen},
	linkcolor={darkred},
	urlcolor={darkblue},
	pdfpagemode={UseOutlines}}
\definecolor{darkgreen}{RGB}{50,190,50}
\definecolor{darkblue}{RGB}{0,0,190}
\definecolor{darkred}{RGB}{238,0,0}
\usepackage{soul}
\newcommand{\ie}{\textit{i.e.}}
\newcommand{\eg}{\textit{e.g.}}

\newcommand{\bra}[1]{\ensuremath{\left\langle\right. #1 \left.\right|}}
\newcommand{\ket}[1]{\ensuremath{\left|\right. #1 \left.\right\rangle}}

\newcommand{\fbra}[1]{\ensuremath{\left\langle\!\left\langle\right.\right.\! #1 \!\left.\left.\right|\hspace*{-0.75pt}\right|}}
\newcommand{\fket}[1]{\ensuremath{\left|\hspace*{-0.75pt}\left|\right.\right.\! #1 \!\left.\left.\right\rangle\!\right\rangle}}

\newcommand{\anticomm}[2]{\ensuremath{\left\{\right.\! #1 \,, #2 \!\left.\right\}}}

\newcommand{\fscpr}[2]{\ensuremath{\left\langle\!\hspace*{-0.7pt}\left\langle\right.\right.\! #1 \!\left.\left.\right|\!\right| #2 \!\left.\left.\right\rangle\!\hspace*{-0.7pt}\right\rangle}}

\newcommand{\pr}{^{\prime}}

\newcommand{\tr}{\textnormal{Tr}}
\newcommand{\djj}{d\kern-0.4em\char"16\kern-0.1em}

\begin{document}

\title{Reasonable fermionic quantum information theories require relativity
%%Symmetric pure state marginals without~a consistent tensor product structure
}
\author{Nicolai Friis}
\email{nicolai.friis@uibk.ac.at}
\affiliation{
Institute for Theoretical Physics, University of Innsbruck,
Technikerstra{\ss}e 21a,
A-6020 Innsbruck,
Austria}
\affiliation{
Institute for Quantum Optics and Quantum Information,
Austrian Academy of Sciences,
Technikerstra{\ss}e 21a,
A-6020 Innsbruck,
Austria}
\date{\today}
\begin{abstract}
%%Abstract 573 characters incl spaces
We show that any quantum information theory based on anticommuting operators must be supplemented by a superselection rule deeply rooted in relativity to establish a reasonable notion of entanglement. While quantum information may be encoded in the fermionic Fock space, the unrestricted theory has a peculiar feature: The marginals of bipartite pure states need not have identical entropies, which leads to an ambiguous definition of entanglement. We solve this problem, by proving that it is removed by relativity, i.e., by the parity superselection rule that arises from Lorentz invariance
via the spin-statistics connection. Our results hence unveil a fundamental conceptual inseparability of quantum information and the causal structure of relativistic field theory.
\end{abstract}
\pacs{
%03.67.Lx,   %Quantum computation architectures and implementations
03.67.Mn, %Entanglement and quantum nonlocality in quantum information,
05.30.-d, %Fermi-Dirac statistics,
%42.50.-p,   %Quantum optics
%37.10.Ty    %Ion trapping
%03.65.Ud,   %Entanglement and quantum non-locality
%85.25.-j    %Superconducting devices
11.10.-z   %Field theory
%\hspace*{1cm}DOI:\ \href{http://dx.doi.org/}{1234}
}

\maketitle
%\newpage
%%

\section{Introduction}\label{sec:introduction}

Over the past two decades, quantum information (QI) theory has been developed into~a rich and highly successful framework for information processing. Operating in the domain of quantum mechanical Hilbert spaces, many QI tasks can be approached from a viewpoint that is strongly influenced by computer science, while the physical systems represented by the Hilbert spaces are sometimes of secondary concern. This abstraction from the physical context is a~significant virtue of QI theory. Dealing with problems purely on the level of~a Hilbert space, its subsystems, and operations thereon, without reference to the specific physical implementation, provides~a level of freedom and generality that is highly desirable. Statements can be made for \emph{all} Hilbert spaces of~a certain type. For instance, quantum information processing with qubits can typically be investigated without reference to their implementation \textemdash\ although examples exist, where specifying the encoding of the qubit is relevant for an abstract problem, see, \eg, Ref.~\cite{FriisDunjkoDuerBriegel2014}. Besides multi-qubit systems, quantum harmonic oscillators are prominent examples for successful abstraction. Based on the commutation relations,~a bosonic Fock space is constructed, that provides a~playground for quantum optics, irrespective of the particular realization, be it as optical modes, superconducting circuits, or vibrational degrees of freedom, to name only~a few examples.

Here, another type of Hilbert space \textemdash\ the fermionic Fock space \textemdash\ will be considered. That is, the basic algebra is based on anticommuting operators, rather than commuting ones. Absent physical interpretation, one may yet work with such~a Hilbert space, identify its subsystems, and their correlations. In other words, one may attempt to construct an abstract fermionic QI theory, see, \eg, Refs.~\cite{SchliemannLossMacDonald2001,SchliemannCiracKusLewensteinLoss2001,LiZengLiuLong2001,EckertSchliemannBrussLewenstein2002,
Shi2003,BoteroReznik2004,CabanPodlaskiRembielinskiSmolinskiWalczak2005,BanulsCiracWolf2007,BalachandranGovindarajanDeQueirozReyesLega2013}. Conceptually, it is of great importance to collect all types of particles encountered in nature in a common framework, and hence strengthen the generality of QI theory as a whole. However, as we shall discuss, the physically unrestricted fermionic QI theory suffers from~a disconcerting malady: As noted already a decade ago~\cite{Moriya2002,ArakiMoriya2003,Moriya2005}, the marginals of bipartite fermionic pure states may not have matching spectra. This leaves the typical notion of entropy of entanglement in~a state of ambiguity due to the mismatch of reduced state entropies. Depending on the choice of subsystem, different amounts of entanglement would be attributed to the system. Indeed, facing a globally pure state, for which one subsystem is maximally mixed, while the other remains pure would be a significant concern, for instance, in connection with Hawking radiation and the black hole information paradox (see, \eg,~\cite{Harlow2016}). These problems do not occur in theories with~a natural tensor product structure, like bosonic modes or qubits, where the Schmidt decomposition guarantees symmetric marginal entropies for pure states. For fermions, on the other hand, mappings to~a tensor product space, \ie, to qubits, do not generally preserve the structure of the subsystems~\cite{FriisLeeBruschi2013}, and the issue persists.

In this work, we resolve this problem. We show that it is overcome by imposing~a superselection rule (SSR) that forbids coherent superpositions of even and odd numbers of fermions. This property of the SSR, which is equivalent to the restriction to parity-conserving operator \mbox{(sub-)}algebras, is most fortunate, since the SSR also gives rise to~a natural definition of subsystems~\cite{BanulsCiracWolf2007}, and is hence widely used. Although, as we shall show, the problem of asymmetric pure state marginals is thus removed, it seems rather artificial to enforce such~a restriction within the abstract theory. In particular, since the often referenced argument by Wick, Wightman and Wigner~\cite{WickWightmanWigner1952} in defense of this SSR is based on time reversal symmetry, which cannot be an exact symmetry of nature in the face of charge-parity (CP) violation~\cite{ChristensonCroninFitchTurlay1964,AlaviHaratiEtAlKTeVCollaboration1999} and the charge-parity-time (CPT) theorem~\cite{Lueders1954,PauliRosenfeldWeisskopf1955,Lueders1957}. In contrast, we discuss the later argument for the SSR from~\cite{HegerfeldtKrausWigner1968} invoking invariance under rotations by $2\pi$. Crucially, we note that the causal structure of relativistic quantum field theory (QFT) enters into this line of reasoning, i.e., Lorentz invariance is required to establish the spin-statistics connection. We argue that, only once the fermionic model is embedded in~a physical context, in this case relativistic QFT, does the SSR arise naturally. The significance of this approach is twofold. First, it ensures that the entropy of entanglement remains a well-defined concept in physical QI theories. Second, it provides a new perspective on the motivation for using the SSR, which has been employed in detailed studies of fermionic entanglement, such as~\cite{BanulsCiracWolf2007}. Third, this result knits together the fabrics of QI and relativistic QFT in a fundamental way: We argue that the fermionic theory, and by extension all quantum information theory, must be viewed in the physical context of relativity.

Besides the possible interest for fermionic QI theory and applications such as entanglement within QFT in curved spacetimes, this work adds~a new facet to the discussion of informational constructions of quantum theory (see, \eg, Refs.~\cite{ChiribellaDArianoPerinotti2010,ChiribellaDArianoPerinotti2011,ChiribellaDArianoPerinotti2012}), by introducing an information-theoretic aspect of SSRs \textemdash\ a~fascinating topic in its own right (see, \eg, Refs.~\cite{WickWightmanWigner1952,AharonovSusskind1967,Yoshihuku1972,StrocchiWightman1974,VerstraeteCirac2003,
BartlettWiseman2003,SchuchVerstraeteCirac2004,SkotioniotisTolouiDurhamSanders2013,SkotioniotisTolouiDurhamSanders2014} for~a selection of literature). It is also of interest to note that the SSR does not remove the intrinsically different character of fermionic modes and qubits, as indicated by the existence of pure states that satisfies the SSR, but still cannot be consistently mapped to multi-qubit states, see the~\hyperref[sec:appendix]{Appendix}.

In the following, we will first outline the construction of the fermionic Fock space, as well as of the pure and mixed states in such~a Hilbert space in Sec.~\ref{sec:framework}. To understand the origin of the problem described above, we will then discuss the subtleties involved in forming subsystems of fermionic modes in Sec.~\ref{sec:partitioning the fermionic fock space}, and give an example for~a pure state that features marginals with different entropies. Finally, the role and the origin of SSRs are discussed in Sec.~\ref{sec:invoking relativity}, and we show how the problem can be disposed of in Sec.~\ref{eq:proof of reasonable fermionic QI}.

\section{Fermionic Fock space}\label{sec:framework}

Let us consider~a system of~$n$ fermionic modes with mode operators~$b_{k}$ and~$b_{k}^{\dagger}$ for~$(k=1,\ldots,n)$, which satisfy the anticommutation relations
\begin{align}
    \anticomm{b_{i}}{b_{j}^{\dagger}}   &=\,\delta_{ij}\,,
    \qquad
    \anticomm{b_{i}}{b_{j}}   \,=\,0\,,
    \label{eq:anticomm relations}
\end{align}
for all~$i,j$. The vacuum state is annihilated by all~$b_{k}$, \ie, $b_{k}\fket{\!0\!}=0\ \forall\,k$, and the purpose of the double-lined notation for the state vectors will become apparent shortly. The creation operators~$b_{k}^{\dagger}$ populate the vacuum with single fermions, that is, $b_{k}^{\dagger}\fket{\!0\!}=\fket{\!1_{k}\!}$. When two, or more, fermions are created, the corresponding tensor product of single-particle states needs to be antisymmetrized due to the indistinguishability of the particles. We use the convention
\begin{align}
    b_{k}^{\dagger}b_{k\pr}^{\dagger}\fket{\!0\!}   &=\,\fket{\!1_{k}\!}\wedge\fket{\!1_{k\pr\!}\!}=\fket{\!1_{k}\!}\fket{\!1_{k\pr\!}\!}\,,
    \label{eq:fermion two particle state}
\end{align}
where we use the double-lined notation to imply the antisymmetrized wedge product ``$\wedge$" between single-mode state vectors with particle content (as opposed to the standard notation $\ket{\cdot}\ket{\cdot}=\ket{\cdot}\otimes\ket{\cdot}$). With this definition at hand, and postponing possible physical restrictions, one may write arbitrary pure states on the Fock space as
\begin{align}
    \fket{\!\Psi\!}    &=
    \gamma_{0}\fket{0}+\sum\limits_{i=1}^{n}\gamma_{i}\fket{\!1_{i}\!}+
    \sum\limits_{j,k}\gamma_{jk}\fket{\!1_{j}\!}\fket{\!1_{k}\!}+\ldots\,.
    \label{eq:fermionic Fock space general state rewritten}
\end{align}
where the complex coefficients~$\gamma_{0}$, $\gamma_{i}$, $\gamma_{jk},\,\ldots$ are chosen such that the state is normalized. Similarly, mixed state density operators can be written as convex sums of projectors on such pure states. For more details on this notation and the fermionic Fock space, see, \eg,~\cite{Friis:PhD2013,FriisLeeBruschi2013}.\\

\section{Partitioning the fermionic Fock space}\label{sec:partitioning the fermionic fock space}

The basic ingredients for an abstract fermionic QI theory are pure states on the total Hilbert space, $\ket{\!\psi\!}\in\mathcal{H}$, and reduced states with respect to some bipartition $(A|B)$, \ie, $\rho_{A(B)}=\tr_{B(A)}\bigl(\ket{\!\psi\!}\!\bra{\!\psi\!}\bigr)$. With this, one may already study correlation measures such as the mutual information, and, most importantly, the entropy of entanglement $\mathcal{E}(\ket{\!\psi\!})=S(\rho_{A})=S(\rho_{B})$, where $S(\rho)=-\tr\bigl(\rho\ln(\rho)\bigr)$. As the first step in such a construction, it is hence necessary to establish~a meaningful notion of subsystems. Since the particle number in the Fock space need not be fixed, we will consider entanglement between different modes. However, due to the antisymmetrization, the Fock space is not naturally equipped with~a tensor product structure with respect to the individual mode subspaces. These subspaces may nonetheless be defined by invoking consistency conditions~\cite{FriisLeeBruschi2013} that ensure that the expectation values of all local observables $\mathcal{O}_{\!A}$ (\ie, as in, operators pertaining only to the modes of the subspace~$A$) yield the same result for the global state~$\rho_{AB}$, and for the corresponding local reduced states~$\rho_{A}=\tr_{B}(\rho_{AB})$, \ie,
\begin{align}
    \left\langle\mathcal{O}_{\!A}\right\rangle_{\rho_{AB}}    &=\,\left\langle\mathcal{O}_{\!A}\right\rangle_{\rho_{A}}\,.
    \label{eq:consistency conditions}
\end{align}
This procedure uniquely defines the mode subspace marginals of any global state, \ie, the partial trace operation in the following way. For an arbitrary number of modes, up to~a scalar prefactor, all density matrix elements of the reduced state $\rho_{A}$ that is obtained by ``tracing out" one mode labelled by $k$ can be brought to the form $b_{\mu_{1}}^{\dagger}\ldots b_{\mu_{i}}^{\dagger}\,\fket{0}\!\fbra{0}\,b_{\nu_{1}}\ldots b_{\nu_{j}}$ for some $\mu_{m}, \nu_{n}\neq k$ where $m=1,\ldots,i$ and $n=1,\ldots,j$. The information about these matrix elements arises from ignoring (or ``forgetting") whether the mode $k$ is occupied or not. That is, the partial trace operation may be defined as
\begin{align}
    &\tr_{k}\bigl(b_{\mu_{1}}^{\dagger}\ldots b_{\mu_{i}}^{\dagger}\,b_{k}^{\dagger}\fket{0}\!\fbra{0}b_{k}\,b_{\nu_{1}}\ldots b_{\nu_{j}}\bigr)\nonumber\\[1mm]
    &\ =\, \tr_{k}\bigl(b_{\mu_{1}}^{\dagger}\ldots b_{\mu_{i}}^{\dagger}\,\fket{0}\!\fbra{0}\,b_{\nu_{1}}\ldots b_{\nu_{j}}\bigr)
    \nonumber\\[1mm]
    &\ =\,b_{\mu_{1}}^{\dagger}\ldots b_{\mu_{i}}^{\dagger}\fket{0}\!\fbra{0}b_{\nu_{1}}\ldots b_{\nu_{j}}\,,
    \label{eq:fermions partial trace}
\end{align}
while all other elements of $\rho_{AB}$, proportional to operators of the form $b_{\mu_{1}}^{\dagger}\ldots b_{\mu_{i}}^{\dagger}\,\fket{0}\!\fbra{0}b_{k}\,b_{\nu_{1}}\ldots b_{\nu_{j}}$ and $b_{\mu_{1}}^{\dagger}\ldots b_{\mu_{i}}^{\dagger}\,b_{k}^{\dagger}\fket{0}\!\fbra{0}\,b_{\nu_{1}}\ldots b_{\nu_{j}}$ do not contribute to the reduced state when $k$ is traced out. The rule can hence be expressed as follows: Operators corresponding to modes that are being traced out are anticommuted towards the vacuum projector before being removed. The ordering of the operators on the left-hand side of Eq.~(\ref{eq:fermions partial trace}), that is, the position of the operators $b_{k}$ and $b_{k}^{\dagger}$ relative to the other annihilation and creation operators, is determined by the condition in Eq.~(\ref{eq:consistency conditions}). Intuitively, it can be seen that the ordering is chosen such that the number of commutations to move $b_{k}$ and $b_{k}^{\dagger}$ towards the projector on the vacuum state keeps track of any signs incurred when applying observables $\mathcal{O}_{\!A}$ to the state $\rho_{AB}$ before the partial trace and commuting these past the operators $b_{k}$ and $b_{k}^{\dagger}$. A more detailed derivation and a rigorous proof of this argument is given in Ref.~\cite{FriisLeeBruschi2013}. An equivalent notion of subsystems may alternatively be obtained by restricting the operator algebras to the corresponding subalgebras for~$A$ and~$B$, see Ref.~\cite{BalachandranGovindarajanDeQueirozReyesLega2013}.

\begin{figure}[ht!]
\label{fig:fermions not qubits}
\includegraphics[width=0.35\textwidth]{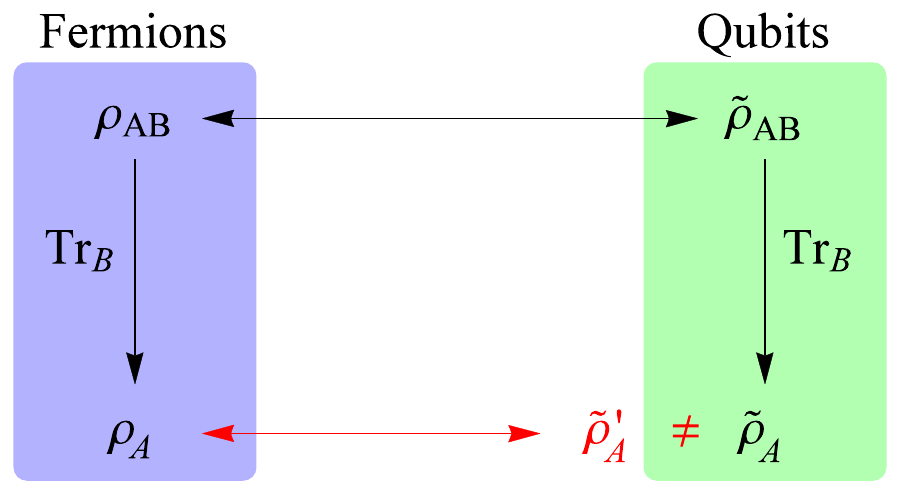}
\caption{%(Color online)
\textbf{Inconsistency between fermions and qubits:} The global $n$-mode fermionic state~$\rho_{AB}$ may be mapped to an isomorphic $n$-qubit state $\tilde{\rho}_{A\!B}$. The marginals of $\rho_{A\!B}$, \eg, $\rho_{A}=\tr_{B}(\rho_{A\!B})$, are well-defined by Eq.~(\ref{eq:consistency conditions}), and may also be mapped to isomorphic qubit states (\eg, $\rho_{A}\leftrightarrow\tilde{\rho}_{A}^{\,\prime}$). However, as shown in Ref.~\cite{FriisLeeBruschi2013}, it is in general impossible to match all the marginals of the $n$-mode state to all marginals of the $n$-qubit state, $\tilde{\rho}_{A}^{\,\prime}\neq\tilde{\rho}_{A}=\tr_{B}(\tilde{\rho}_{A\!B})$. An example for~a state featuring this problem can be found in the~\hyperref[sec:appendix]{Appendix}.}
\end{figure}

At this point, it is helpful to understand the differences between fermionic modes and qubits. For any fixed number~$n$, the fermionic $n$-mode Fock space is isomorphic to an $n$-qubit space. A~widely known example for such an isomorphism is the Jordan-Wigner transformation, (see, \eg,~\cite{BanulsCiracWolf2007}). Such mappings generally do not commute with the procedure of partial tracing~\cite{FriisLeeBruschi2013}, since local mode operators are mapped to global qubit operations. In other words, it is generally not possible to establish isomorphisms between~a fermionic $n$-mode state and an $n$-qubit state in such~a way that also all of the respective fermionic marginals are isomorphic to their qubit counterparts. An illustration of this problem is shown in Fig.~\ref{fig:fermions not qubits}. Consequently, the (quantum) correlations between~$n$ fermionic modes may generally not be identified with those of the isomorphic $n$-qubit states.

In spite of this inequivalence, the partial trace, and hence the subsystems and their entropies remain well defined for fermionic modes. Moreover, the construction of the density operators and its marginals is based solely on the algebraic structure of~(\ref{eq:anticomm relations}), together with the requirement that the expectation values of subsystem observables yield the same result when evaluated using either the global states or the corresponding marginals. No other assumptions are required for~a consistent definition of the subsystems, and their total correlations. For instance, the mutual information~$\mathcal{I}_{AB}$, a~measure of the overall correlation between subsystems~$A$ and~$B$, is in this context already well defined by the expression
\begin{align}
    \mathcal{I}_{AB}(\rho_{AB}) &=\,S(\rho_{A})\,+\,S(\rho_{B})\,-\,S(\rho_{AB})\,,
    \label{eq:mutual inf definition}
\end{align}
where~$\rho_{A(B)}=\tr_{B(A)}(\rho_{AB})$, and~$S(\rho)=-\tr\bigl(\rho\ln(\rho)\bigr)$ is the von~Neumann entropy of the density operator~$\rho$. But, as we shall elaborate on shortly, the same cannot be said for \emph{genuine} quantum correlations, \ie, entanglement. Consider, for instance, the non-superselected two-mode pure state given by
\begin{align}
    \fket{\!\psi\!} &=\gamma_{0}\fket{\!0\!}+\gamma_{k}\fket{\!1_{k}\!}+\gamma_{k\pr}\fket{\!1_{k\pr\!}\!}+\gamma_{kk\pr}\fket{\!1_{k}\!}\!\fket{\!1_{k\pr\!}\!}.
    \label{eq:two-mode non-superselected state}
\end{align}
According to the prescription of Eq.~(\ref{eq:fermions partial trace}), the single-mode reduced states can be quickly checked to be
\begin{subequations}
\label{eq:single mode reduction no superselection}
\begin{align}
    \rho_{k}    &=\tr_{k\pr}\bigl(\fket{\!\psi\!}\!\fbra{\!\psi\!}\bigr)=\bigl(|\gamma_{0}|^{2}+|\gamma_{k\pr}|^{2}\bigr)\fket{\!0\!}\!\fbra{\!0\!}
    \nonumber\\[1mm]
    &\ +\bigl(|\gamma_{k}|^{2}+|\gamma_{kk\pr}|^{2}\bigr)\fket{\!1_{k}\!}\!\fbra{\!1_{k}\!}\label{eq:single mode reduction no superselection k}\\[1mm]
    &\ +\Bigl[\bigl(\gamma_{0}\gamma_{k}^{*}+\gamma_{k\pr}\gamma_{kk\pr}^{*}\bigr)\fket{\!0\!}\!\fbra{\!1_{k}\!}+\mathrm{H.~c.}\Bigr],\nonumber\\[1mm]
    \rho_{k\pr}    &=\tr_{k}\bigl(\fket{\!\psi\!}\!\fbra{\!\psi\!}\bigr)=\bigl(|\gamma_{0}|^{2}+|\gamma_{k}|^{2}\bigr)\fket{\!0\!}\!\fbra{\!0\!}
    \nonumber\\[1mm]
    &\ +\bigl(|\gamma_{k\pr}|^{2}+|\gamma_{kk\pr}|^{2}\bigr)\fket{\!1_{k\pr\!}\!}\!\fbra{\!1_{k\pr\!}\!}\label{eq:single mode reduction no superselection kpr}\\[1mm]
    &\ +\Bigl[\bigl(\gamma_{0}\gamma_{k\pr}^{*}-\gamma_{k}\gamma_{kk\pr}^{*}\bigr)\fket{\!0\!}\!\fbra{\!1_{k\pr\!}\!}+\mathrm{H.~c.}\Bigr],\nonumber
\end{align}
\end{subequations}
where the symmetry between the subsystems is broken by the different relative signs within the off-diagonal elements. The eigenvalues of the two reduced states do not match in general. For example, when $\gamma_{0}=\gamma_{k}=\gamma_{k\pr}=\gamma_{kk\pr}=1/2$, the mode~$k$ appears to be in~a pure state (with eigenvalues $0$ and $1$), whereas the state of the mode~$k\pr$ is maximally mixed (both eigenvalues are $1/2$). Normally, the entropy of the subsystem of~a pure state would be considered as~an entanglement measure. Here, depending on the choice of subsystem, one would either conclude that the overall state is maximally entangled, or not entangled at all. This problem is not limited to pure states. It persists for mixed states, where the entropy of entanglement is of central importance for the entanglement of formation. Such~an ambiguity in the definition of entanglement is of course highly undesirable. One possibility to resolve the issue, would be to change the definition of entanglement, and work with~a non-symmetric quantity. On the other hand, such~a drastic step may not be required, if one is willing to embed the abstract fermionic quantum information theory in~a physical framework. As we shall show in the following,~a reasonable definition of entanglement between fermionic modes is obtained when invoking an additional physical principle\textemdash the spin-statistics connection, which itself arises from (special) relativity.

\section{Invoking relativity\textemdash superselection rules}\label{sec:invoking relativity}

No SSRs have been introduced up to this point. Note that the term SSR may refer to different restrictions. For instance, they may arise from fundamental symmetries of the system, such as parity~\cite{WickWightmanWigner1952}, or charge conservation~\cite{AharonovSusskind1967,Yoshihuku1972,StrocchiWightman1974}. Alternatively, effective (or generalized) SSRs originate from practical limitations, such as particle number conservation due to energy constraints, see, \eg,~\cite{VerstraeteCirac2003,BartlettWiseman2003,WisemanVaccaro2003,WisemanBartlettVaccaro2004,SchuchVerstraeteCirac2004}. Both type of restrictions may be formalized as constraints on the observables, see, \eg,~\cite{BoteroReznik2004,CabanPodlaskiRembielinskiSmolinskiWalczak2005}.

Here, we will formulate such constraints in~a different, but equivalent way, as restrictions on the components~$\gamma_{0}$, $\gamma_{i}$, $\gamma_{jk},\,\ldots\,$ [see Eq.~(\ref{eq:fermionic Fock space general state rewritten})], of pure state decompositions with respect to the Fock basis. In particular, we will consider any coherent superpositions of even and odd numbers of fermionic excitations as unphysical. The argument that we will employ to defend this position is based on the well-known spin-statistics connection, relating the anticommutation relation algebra to the transformation properties associated to objects of half-integer spin. Recall that, \emph{a priori}, we have made no assumptions on the physical realization of the anticommutation relations of Eq.~(\ref{eq:anticomm relations}). Nonetheless, as an inevitable consequence of this anticommutation relation algebra, the excitations of the mode operators must satisfy Fermi-Dirac statistics.

To explain this, we must surrender some level of abstraction and provide~a physical context. When we place the fermionic quantum information theory in the context of relativistic field theory, the spin-statistics connection follows from Lorentz invariance (see, \eg, Refs.~\cite{Pauli1940,Schwinger1951} and~\cite[pp.~52]{PeskinSchroeder1995}). Thus, the (special) theory of relativity enforces that we must interpret the fermionic excitations of our theory as particles of half-integer spin. In other words, relativity establishes a connection between the algebra of the mode operators and the transformation properties under spatial rotations of the corresponding fermionic particles. In particular, the states of odd numbers of these fermions switch their sign under rotations by~$2\pi$, while even numbers of fermions are left invariant. If superpositions of even and odd numbers of fermions were permitted, rotations by~$2\pi$ would change the relative sign of the respective contributions within the superposition. For example, suppose that an observer was able to prepare~a superposition $\fket{\!+\!}=\bigl(\fket{0}+\fket{1_{k}}\bigr)/\sqrt{2}$ in some fermionic mode~$k$. By rotating their frame by~$2\pi$, the state $\fket{\!+\!}$ is converted to the orthogonal state $\fket{\!-\!}=\bigl(\fket{0}-\fket{1_{k}}\bigr)/\sqrt{2}$. Here, one may think of the observer physically rotating their preparation device relative to the detector distinguishing the orthogonal states $\fket{\!+\!}$ and $\fket{\!-\!}$. In other words, if one demands that rotations by $2\pi$ should leave physics unchanged, one thereby enforces the parity SSR that forbids superpositions of even and odd numbers of fermions~\cite{HegerfeldtKrausWigner1968}.

Crucially, this line of reasoning is intimately tied to relativity, whereas the argument for the parity SSR presented in Ref.~\cite{WickWightmanWigner1952} is based\footnote{Indeed, Ref.~\cite{WickWightmanWigner1952} hints at the possibility of basing the proof on rotations instead of time-inversion, but this argument is not elaborated on in~\cite{WickWightmanWigner1952} and was only later published in~\cite{HegerfeldtKrausWigner1968}.} on time-inversion symmetry. However, with the observation of charge-parity (CP) violation~\cite{ChristensonCroninFitchTurlay1964,AlaviHaratiEtAlKTeVCollaboration1999}, the charge-parity-time (CPT) theorem~\cite{Lueders1954,PauliRosenfeldWeisskopf1955,Lueders1957} suggests that time-reversal is no (exact) symmetry either. Here, on the other hand, the argument for the parity SSR is constructed such that we rely only on Lorentz invariance via the spin-statistics connection, which happens to also be a central requirement for CPT symmetry. On the other hand, the SSR constraint imposed in this way provides the essential ingredient to guarantee~a meaningful quantum (information) theory based on anticommuting operators, as we will show next.

\section{Superselection rules \& symmetric pure state marginals}\label{eq:proof of reasonable fermionic QI}

We shall now provide the main technical statement of this work, and its proof.\\

\vspace*{-1mm}
%\begin{theorem}
\textbf{Theorem}. The marginals $\rho_{A(B)}=\tr_{B(A)}\bigl(\fket{\!\psi\!}\!\fbra{\!\psi\!}\bigr)$ of any bipartition $(A|B)$ of~a pure state $\fket{\!\psi\!}$ in the fermionic Fock space have the same spectrum, if $\fket{\!\psi\!}$ satisfies the SSR prohibiting superpositions of even and odd numbers of fermions, \ie,
\begin{align*}
    \fket{\!\psi\!}\ \, \mbox{satifies SSR}\, \ \Rightarrow\, \ \operatorname{spetr}(\rho_{A})=\operatorname{spetr}(\rho_{B})\ \forall \ (A|B).
\end{align*}
%\end{theorem}
\begin{proof}
To show this, let us consider~a pure state~$\fket{\!\psi_{\mathrm{even}}^{N}\!}$ in an~$n$-mode fermionic Fock space, where $N=\{1,2,\ldots,n\}$ denotes the set of modes, and without loss of generality we have chosen~$\fket{\!\psi_{\mathrm{even}}^{N}\!}$ to be~a superposition of states with even numbers of fermions. The set~$N$ is then partitioned into the subsets $M=\{\mu_{i}|\,\mu_{i,j}\in N: \mu_{i}\neq\mu_{j}\,\mbox{if}\,i\neq j; i,j=1,2,\ldots,m<n\}$ and $M^{C}=N\backslash M$, such that $N=M\cup M^{C}$ and $M\cap M^{C}=\emptyset$. With respect to this bipartition, we may write the state~$\fket{\!\psi_{\mathrm{even}}^{N}\!}$ in~the pure state decomposition
\begin{align}
    \fket{\!\psi_{\mathrm{even}}^{N}\!} &=\,\gamma_{0}\fket{\!0\!}\,+\,\sum\limits_{i=1}^{m}\gamma_{\mu_{i}}\fket{\!1_{\mu_{i}}\!}\!\fket{\!\psi^{M^{C}}_{\mu_{i},\mathrm{odd}}\!}
    \label{eq:proof ansatz}\\%[1mm]
    &\ +\,\sum\limits_{\substack{i,j=1\\ j>i}}^{m}\gamma_{\mu_{i}\mu_{j}}\fket{\!1_{\mu_{i}}\!}\!\fket{\!1_{\mu_{j}}\!}\!\fket{\!\psi^{M^{C}}_{\mu_{i}\mu_{j},\mathrm{even}}\!}\,+\,\ldots
    \nonumber\\%[1mm]
    &\ +\,\gamma_{\mu_{1}\ldots\mu_{m}}\fket{\!1_{\mu_{1}}\!}\ldots\fket{\!1_{\mu_{m}}\!}\!\fket{\!\psi^{M^{C}}_{\mu_{1}\ldots\mu_{m},\mathrm{odd}}\!}\,,
    \nonumber
\end{align}
where $\gamma_{\mu_{i}},\ldots,\gamma_{\mu_{1}\ldots\mu_{m}}\in\mathds{C}$, $\fscpr{\!\psi_{\mathrm{even}}^{N}\!}{\!\psi_{\mathrm{even}}^{N}\!}=1$, and without loss of generality we have here selected~$m$ to be odd.
The states $\fket{\!\psi^{M^{C}}_{\mu_{i}\ldots\mu_{k},\mathrm{even (odd)}}\!}$ contain only even (odd) numbers of excitations, and only in modes from the set~$M^{C}$. Any sign changes that may occur when rewriting~a given state in such~a decomposition can be absorbed into the $\gamma$-coefficients. Adhering to the ``outside-in" tracing rule of Eq.~(\ref{eq:fermions partial trace}), we note that the state has been brought to~a form, where the partial trace over~$M^{C}$ is achieved by simply removing all projectors $\fket{\!\psi^{M^{C}}_{\mu_{i}\ldots\mu_{k}}\!}\!\fbra{\!\psi^{M^{C}}_{\mu_{i}\ldots\mu_{k}}\!}$ pertaining to~$M^{C}$ from the projector on $\fket{\!\psi_{\mathrm{even}}^{N}\!}$, without incurring any additional sign flips. On the other hand, if we trace over the modes in the set~$M$ instead, anticommuting the operators corresponding to modes in~$M$ towards the vacuum projector in the process, we may generate sign changes. However, for the superselected state, all the nonzero contributions to the partial trace over~$M$ are generated from elements such as
\begin{align}
    \tr_{M}\bigl(|\gamma_{\mu_{i}}|^{2}\,b^{\dagger}_{\mu_{i}}\fket{\!\psi^{M^{C}}_{\mu_{i},\mathrm{odd}}\!}\!\fbra{\!\psi^{M^{C}}_{\mu_{i},\mathrm{odd}}\!}\,b_{\mu_{i}}\bigr)\nonumber\\
    =\,|\gamma_{\mu_{i}}|^{2}\,\fket{\!\psi^{M^{C}}_{\mu_{i},\mathrm{odd}}\!}\!\fbra{\!\psi^{M^{C}}_{\mu_{i},\mathrm{odd}}\!}\,.
\end{align}
There, the parity of the number of anticommutations towards $\fket{\!0\!}$ from the left, is the same as the parity of the number of anticommutations towards $\fbra{\!0\!}$ from the right. In other words, once the state has been brought to the form of Eq.~(\ref{eq:proof ansatz}), the partial trace can be carried out as if operating on~a tensor product of Hilbert spaces, \ie, as if $\mathcal{H}_{N}=\mathcal{H}_{M}\otimes\mathcal{H}_{M^{C}}$. In particular, this implies that the two reduced states on $\mathcal{H}_{M}$ and $\mathcal{H}_{M^{C}}$, which are isomorphic to the corresponding $m$-mode and $(n-m)$-mode reduced fermionic states, respectively, have the same spectrum. For any bipartition of the Fock space,~a decomposition with this property can be found, although, in general, no decomposition exists that simultaneously accomplishes the required task for all bipartitions at once. An example is presented in the \hyperref[sec:appendix]{Appendix}. We hence conclude that the SSR forbidding superpositions of even and odd numbers of fermions, guarantees that the spectra of the marginals for any bipartition are pairwise identical. An analogous argument applies if the initial state is~a superposition of only odd numbers of fermions, or if~$m$ is even, and the proof therefore applies without restriction.
\end{proof}

\vspace*{5mm}

\section{Discussion}\label{sec:conclusion}

Within quantum information theory and quantum computation, discussing problems in an abstract context has proven to be very useful. However, when attempting~a similar approach to~a quantum information theory based on~a fermionic Fock space one encounters difficulties. As we have shown, an unrestricted fermionic model features pure states with non-symmetric bipartitions. That is, pairs of reduced states across bipartitions need not have the same spectra, which is very problematic for the definition of entanglement. As we have shown, this problem is removed, when superpositions of even and odd numbers of fermions are forbidden. Nonetheless, the removal of an inconvenience appears to be~a rather weak justification for the introduction of~a restriction of generality within the abstract model. On the other hand, when placing the fermionic system within the physical framework of~a relativistic quantum field theory, the SSR follows naturally from the requirements of Lorentz invariance and symmetry under rotations by $2\pi$.\\

We hence argue that, in contrast to bosonic or qudit-based variants, any fermionic quantum information theory \emph{must} be seen as (part of)~a relativistic quantum field theory. Without such a physical context, a reasonable quantum information theory based on anticommuting operators can be obtained by adding the parity SSR as an axiom, but such an approach would seem to be rather \emph{ad hoc}. This strong hint at the inseparability of quantum information theory and the theory of relativity is rather surprising, but may provide deeper insight into constructions of quantum theory based on informational principles, see, \eg, Refs.~\cite{ChiribellaDArianoPerinotti2010,ChiribellaDArianoPerinotti2011,ChiribellaDArianoPerinotti2012}. Moreover, our results provide fresh insight into the debate of entanglement in systems of indistinguishable particles (see, \eg,~\cite{WisemanVaccaro2003,WisemanBartlettVaccaro2004}) in general, and questions of entanglement between fermionic modes~\cite{EislerZimboras2015} specifically. Finally, it will be of significant interest to see to what extent simulations of fermions, for instance in superconducting materials~\cite{Barends-Martinis2015}, or in graphene~\cite{IorioLambiase2013,Iorio2015}, can capture the behaviour of superpositions of different particle numbers.\\[2mm]

%%
%\vspace*{-10mm}
%%
\begin{acknowledgements}
We are grateful to Sergey Filippov for valuable discussions, and for bringing the problem of non-symmetric pure state marginals to our attention. We are also grateful to Vedran Dunjko, Geza Giedke, Michalis Skotiniotis, and Sterling Archer for their enlightening comments, and to the referees of the New Journal of Physics for their help in correcting an error in an earlier version of the manuscript. This work has been supported in part by the Austrian Science Fund (FWF) through the SFB FoQuS: F4012.
\end{acknowledgements}

\newpage
\appendix*
%\section*{Appendix}
%\renewcommand\appendixname{}
%\renewcommand{\thesection}{A.\arabic{section}}

\section{Superselected fermionic pure state inequivalent to qubits state}\label{sec:appendix}

Here, we present an example for~a pure state of four fermionic modes that satisfies the superselection rule (SSR) that forbids superpositions of even and odd numbers of fermions. We explicitly compute the marginals of this state and show that the subsystem spectra match for any bipartition. We further prove that, nonetheless, this state and its marginals do not admit~a consistent mapping to~a four-qubit state (and its marginals). The most general pure state for four fermionic modes, labelled $1,2,3,4$, that only contains even numbers of excitations, may be written as
\begin{align}
    \fket{\!\psi_{{\rm even}}^{(4)}\!}  &=\,\alpha_{0}\fket{\!0\!}\,+\,\alpha_{12}\fket{\!1_{1}\!}\!\fket{\!1_{2}\!}\,+\,\alpha_{13}\fket{\!1_{1}\!}\!\fket{\!1_{3}\!}
    \nonumber\\[1mm]
    &\ \,+\,\alpha_{14}\fket{\!1_{1}\!}\!\fket{\!1_{4}\!}\,+\,\alpha_{23}\fket{\!1_{2}\!}\!\fket{\!1_{3}\!}
    \nonumber\\[1mm]
    &\ \,+\,\alpha_{24}\fket{\!1_{2}\!}\!\fket{\!1_{4}\!}\,+\,\alpha_{34}\fket{\!1_{3}\!}\!\fket{\!1_{4}\!}
    \nonumber\\[1mm]
    &\ \,+\,\alpha_{1234}\fket{\!1_{1}\!}\!\fket{\!1_{2}\!}\!\fket{\!1_{3}\!}\!\fket{\!1_{4}\!}\,,
    \label{eq:four mode even pure state}
\end{align}
where $|\alpha_{0}|^{2}+|\alpha_{12}|^{2}+|\alpha_{13}|^{2}+|\alpha_{14}|^{2}+|\alpha_{23}|^{2}+|\alpha_{24}|^{2}+|\alpha_{34}|^{2}+|\alpha_{1234}|^{2}=1$. First, let us consider the bipartitions into two subsets containing one and three modes, respectively. For instance, let us consider the bipartition $(1|2,3,4)$. We may decompose the state~$\fket{\!\psi_{{\rm even}}^{(4)}\!}$ into terms containing excitations in mode~$1$, and those which do not contain such terms, \ie, we write
\begin{align}
    \fket{\!\psi_{{\rm even}}^{(4)}\!}  &=\,\gamma_{1}\,\fket{\!\phi_{1}\!}\,+\,\gamma_{\lnot1}\,\fket{\!\phi_{\lnot1}\!}\,,
    \label{eq:four mode even pure state decomp wrt mode one excitations}
\end{align}
where $|\gamma_{1}|^{2}+|\gamma_{\lnot1}|^{2}=1$, $b_{1}^{\dagger}b_{1}\fket{\!\phi_{1}\!}=\fket{\!\phi_{1}\!}$, and $b_{1}\fket{\!\phi_{\lnot1}\!}=0$. Since $\fscpr{\!\phi_{1}\!}{\!\phi_{\lnot1}\!}=0$, the marginals with respect to the bipartition $(1|2,3,4)$ are then easily obtained as
\begin{subequations}
\label{eq:bipartition 1vs234 marginals}
\begin{align}
    \rho_{1}    &=\,\tr_{2,3,4}\bigl(\fket{\!\psi_{{\rm even}}^{(4)}\!}\!\fbra{\!\psi_{{\rm even}}^{(4)}\!}\bigr)
    \label{eq:bipartition 1vs234 rho 1}\\[1mm]
    &=\,|\gamma_{1}|^{2}\fket{\!1_{1}\!}\!\fbra{\!1_{1}\!}\,+\,|\gamma_{\lnot1}|^{2}\fket{\!0\!}\!\fbra{\!0\!}\,,
    \nonumber\\[1.5mm]
    \rho_{2,3,4}    &=\,\tr_{1}\bigl(\fket{\!\psi_{{\rm even}}^{(4)}\!}\!\fbra{\!\psi_{{\rm even}}^{(4)}\!}\bigr)
    \label{eq:bipartition 1vs234 rho 234}\\[1mm]
    &=\,|\gamma_{1}|^{2}\fket{\!\tilde{\phi}_{1}\!}\!\fbra{\!\tilde{\phi}_{1}\!}\,+\,|\gamma_{\lnot1}|^{2}\fket{\!\phi_{\lnot1}\!}\!\fbra{\!\phi_{\lnot1}\!}\,,
    \nonumber
\end{align}
\end{subequations}
where $\fket{\!\tilde{\phi}_{1}\!}\!\fbra{\!\tilde{\phi}_{1}\!}=\tr_{1}\bigl(\fket{\!\phi_{1}\!}\!\fbra{\!\phi_{1}\!}\bigr)$. Because~$\fket{\!\psi_{{\rm even}}^{(4)}\!}$ contains only even numbers of fermions, the same is true for $\fket{\!\phi_{1}\!}$ and~$\fket{\!\phi_{\lnot1}\!}$, whereas $\fket{\!\tilde{\phi}_{1}\!}$ contains only odd numbers of fermions. Hence we may conclude that $\fscpr{\!\tilde{\phi}_{1}\!}{\!\phi_{\lnot1}\!}=0$, and it is thus easy to see that the marginals have the same spectrum. The same argument goes through for the bipartitions $(2|1,3,4)$, $(3|1,2,,4)$ and $(4|1,2,3)$.\\

Let us now turn to the bipartitions into pairs of modes. We use the tracing rule
\begin{align}
    \tr_{k}\bigl(b_{\mu_{1}}^{\dagger}\ldots b_{\mu_{i}}^{\dagger}\,b_{k}^{\dagger}\fket{0}\!\fbra{0}b_{k}\,b_{\nu_{1}}\ldots b_{\nu_{j}}\bigr)\nonumber\\[1mm]
    =\ b_{\mu_{1}}^{\dagger}\ldots b_{\mu_{i}}^{\dagger}\fket{0}\!\fbra{0}b_{\nu_{1}}\ldots b_{\nu_{j}}\,,
    \label{eq:fermions partial trace appendix}
\end{align}
\ie, operators pertaining to modes that are being traced over are anticommuted towards the projector on the vacuum state, before being removed. This prescription, which uniquely determines the marginals (see, \eg, Ref.~\cite{FriisLeeBruschi2013}), is~a direct consequence of the requirement that expectation values of local observables give the same result when evaluated for the global state, or for the reduced state, that is,
\begin{align}
    \left\langle\mathcal{O}_{\!A}\right\rangle_{\rho_{AB}}    &=\,\left\langle\mathcal{O}_{\!A}\right\rangle_{\rho_{A}}\,.
    \label{eq:consistency conditions appendix}
\end{align}
For the marginals of the bipartition~$(1,2|3,4)$ we find
\begin{subequations}
\label{eq:bipartition 12vs34 marginals}
\begin{align}
    \rho_{1,2}    &=\tr_{3,4}\bigl(\fket{\!\psi_{{\rm even}}^{(4)}\!}\!\fbra{\!\psi_{{\rm even}}^{(4)}\!}\bigr)
    =p_{1,2}^{\mathrm{even}}\,\rho_{1,2}^{\mathrm{even}}+p_{1,2}^{\mathrm{odd}}\,\rho_{1,2}^{\mathrm{odd}}\,,
    \label{eq:bipartition 12vs34 rho 12}\\[1.5mm]
    \rho_{3,4}    &=\tr_{1,2}\bigl(\fket{\!\psi_{{\rm even}}^{(4)}\!}\!\fbra{\!\psi_{{\rm even}}^{(4)}\!}\bigr)
    =p_{3,4}^{\mathrm{even}}\,\rho_{3,4}^{\mathrm{even}}+p_{3,4}^{\mathrm{odd}}\,\rho_{3,4}^{\mathrm{odd}}\,,
    \label{eq:bipartition 12vs34 rho 34}
\end{align}
\end{subequations}
where the reduced state density operators in the even and odd subspaces are given by
\begin{subequations}
\label{eq:bipartition 12vs34 marginals 12}
\begin{align}
    p_{1,2}^{\mathrm{even}}\,\rho_{1,2}^{\mathrm{even}}    &=\,\bigl(|\alpha_{0}|^{2}+|\alpha_{34}|^{2}\bigr)\,\fket{\!0\!}\!\fbra{\!0\!}
    \label{eq:bipartition 12vs34 rho 12 even subspace}\\[1mm]
    &\ +\,\bigl(|\alpha_{12}|^{2}+|\alpha_{1234}|^{2}\bigr)\,\fket{\!1_{1}\!}\!\fket{\!1_{2}\!}\!\fbra{\!1_{2}\!}\!\fbra{\!1_{1}\!}
    \nonumber\\[1mm]
    &\ +\,\Bigl[\bigl(\alpha_{0}\alpha_{12}^{*}+\alpha_{34}\alpha_{1234}^{*}\bigr)\fket{\!0\!}\!\fbra{\!1_{2}\!}\!\fbra{\!1_{1}\!}+\mathrm{H.c.}\Bigr],
    \nonumber\\[1.5mm]
    p_{1,2}^{\mathrm{odd}}\,\rho_{1,2}^{\mathrm{odd}}    &=\,\bigl(|\alpha_{13}|^{2}+|\alpha_{14}|^{2}\bigr)\,\fket{\!1_{1}\!}\!\fbra{\!1_{1}\!}
    \label{eq:bipartition 12vs34 rho 12 odd subspace}\\[1mm]
    &\ +\,\bigl(|\alpha_{23}|^{2}+|\alpha_{24}|^{2}\bigr)\,\fket{\!1_{2}\!}\!\fbra{\!1_{2}\!}
    \nonumber\\[1mm]
    &\ +\,\Bigl[\bigl(\alpha_{13}\alpha_{23}^{*}+\alpha_{14}\alpha_{24}^{*}\bigr)\fket{\!1_{1}\!}\!\fbra{\!1_{2}\!}+\mathrm{H.c.}\Bigr],
    \nonumber
\end{align}
\end{subequations}
for the subspace of modes~$1$ and~$2$. Similarly, for~$3$ and~$4$ we obtain
\begin{subequations}
\label{eq:bipartition 12vs34 marginals 34}
\begin{align}
    p_{3,4}^{\mathrm{even}}\,\rho_{3,4}^{\mathrm{even}}    &=\,\bigl(|\alpha_{0}|^{2}+|\alpha_{12}|^{2}\bigr)\,\fket{\!0\!}\!\fbra{\!0\!}
    \label{eq:bipartition 12vs34 rho 34 even subspace}\\[1mm]
    &\ +\,\bigl(|\alpha_{34}|^{2}+|\alpha_{1234}|^{2}\bigr)\,\fket{\!1_{3}\!}\!\fket{\!1_{4}\!}\!\fbra{\!1_{4}\!}\!\fbra{\!1_{3}\!}
    \nonumber\\[1mm]
    &\ +\,\Bigl[\bigl(\alpha_{0}\alpha_{34}^{*}+\alpha_{12}\alpha_{1234}^{*}\bigr)\fket{\!0\!}\!\fbra{\!1_{4}\!}\!\fbra{\!1_{3}\!}+\mathrm{H.c.}\Bigr],
    \nonumber\\[1.5mm]
    p_{3,4}^{\mathrm{odd}}\,\rho_{3,4}^{\mathrm{odd}}    &=\,\bigl(|\alpha_{13}|^{2}+|\alpha_{23}|^{2}\bigr)\,\fket{\!1_{3}\!}\!\fbra{\!1_{3}\!}
    \label{eq:bipartition 12vs34 rho 34 odd subspace}\\[1mm]
    &\ +\,\bigl(|\alpha_{14}|^{2}+|\alpha_{24}|^{2}\bigr)\,\fket{\!1_{4}\!}\!\fbra{\!1_{4}\!}
    \nonumber\\[1mm]
    &\ +\,\Bigl[\bigl(\alpha_{13}\alpha_{14}^{*}+\alpha_{23}\alpha_{24}^{*}\bigr)\fket{\!1_{3}\!}\!\fbra{\!1_{4}\!}+\mathrm{H.c.}\Bigr].
    \nonumber
\end{align}
\end{subequations}
For the superselected state the even and odd subspaces decouple. We may therefore compare the characteristic polynomials for the even and odd subspaces separately. A~simple computation reveals that both~$p_{1,2}^{\mathrm{even}}\,\rho_{1,2}^{\mathrm{even}}$ and $p_{3,4}^{\mathrm{even}}\,\rho_{3,4}^{\mathrm{even}}$ yield the characteristic polynomial
\begin{align}
    &\det\bigl(p_{1,2}^{\mathrm{even}}\,\rho_{1,2}^{\mathrm{even}}-\lambda\mathds{1}\bigr)\,=\,
    \det\bigl(p_{3,4}^{\mathrm{even}}\,\rho_{3,4}^{\mathrm{even}}-\lambda\mathds{1}\bigr)
    \nonumber\\[1.5mm]
    &=\,\lambda^{2}\,-\,\lambda\bigl(|\alpha_{0}|^{2}+|\alpha_{12}|^{2}+|\alpha_{34}|^{2}+|\alpha_{1234}|^{2}\bigr)
    \nonumber\\[1mm]
    &\ +\,(\alpha_{0}\alpha_{1234})(\alpha_{0}\alpha_{1234})^{*}+(\alpha_{12}\alpha_{34})(\alpha_{12}\alpha_{34})^{*}
    \nonumber\\[1mm]
    &\ -\,(\alpha_{0}\alpha_{1234})(\alpha_{12}\alpha_{34})^{*}-(\alpha_{0}\alpha_{1234})^{*}(\alpha_{12}\alpha_{34})\,.
    \label{eq:bipartition 12vs34 char poly even}
\end{align}
Similarly, for the odd subspace we find
\begin{align}
    &\det\bigl(p_{1,2}^{\mathrm{odd}}\,\rho_{1,2}^{\mathrm{odd}}-\lambda\mathds{1}\bigr)\,=\,
    \det\bigl(p_{3,4}^{\mathrm{odd}}\,\rho_{3,4}^{\mathrm{odd}}-\lambda\mathds{1}\bigr)
    \nonumber\\[1.5mm]
    &=\,\lambda^{2}\,-\,\lambda\bigl(|\alpha_{13}|^{2}+|\alpha_{14}|^{2}+|\alpha_{23}|^{2}+|\alpha_{24}|^{2}\bigr)
    \nonumber\\[1mm]
    &\ +\,(\alpha_{13}\alpha_{24})(\alpha_{13}\alpha_{24})^{*}+(\alpha_{14}\alpha_{23})(\alpha_{14}\alpha_{23})^{*}
    \nonumber\\[1mm]
    &\ -\,(\alpha_{13}\alpha_{24})(\alpha_{14}\alpha_{23})^{*}-(\alpha_{13}\alpha_{24})^{*}(\alpha_{14}\alpha_{23})\,.
    \label{eq:bipartition 12vs34 char poly even}
\end{align}
We hence find that the eigenvalues of the marginals~$\rho_{1,2}$ and~$\rho_{3,4}$ coincide. Some straightforward computation along the same lines confirm that this is also the case for the bipartitions $(1,3|2,4)$ and $(1,4|2,3)$. Now, the interesting aspect of this insight pertains to the fact that the four-mode system may not be consistently mapped to~a four-qubit state. For the latter, the matching subsystem spectra would be guaranteed by the Schmidt decomposition. Recall that the consistency conditions (see Ref.~\cite{FriisLeeBruschi2013}) for partial tracing demand that the numerical value of the expectation values of any ``local" operator (in the sense of mode-subspaces) is independent of being evaluated for the overall state, or for the corresponding local reduced state. These consistency conditions then fix the relative signs of different contributions from matrix elements of the total state to the matrix elements of the reduced states. For the example state at hand, the resulting off-diagonal matrix elements of all two-mode vs. two-mode bipartitions are collected in Table~\ref{table:example state off diags}.

Now, one wonders, whether the pure state~$\fket{\!\psi_{{\rm even}}^{(4)}\!}$ and its marginals can be faithfully represented as~a four-qubit state $\ket{\!\psi_{{\rm even}}^{(4)}\!}\in(\mathbb{C}^{2})^{\otimes4}$. Here, we will call such~a mapping faithful, if all the diagonal matrix elements of $\ket{\!\psi_{{\rm even}}^{(4)}\!}\!\bra{\!\psi_{{\rm even}}^{(4)}\!}$ and its marginals with respect to the computational basis match the diagonal elements of $\fket{\!\psi_{{\rm even}}^{(4)}\!}\!\fbra{\!\psi_{{\rm even}}^{(4)}\!}$ with respect to the Fock basis. For the off-diagonal elements we impose~a slightly weaker condition, \ie, that the absolute values of the off-diagonals (with respect to the respective bases) match. This corresponds to demanding that measurements in the Fock basis are reproduced, and that the marginals have the same spectra. These conditions imply that~a faithful mapping from the Fock basis of four fermionic modes to the computational basis of four qubits must be of the form
\begin{subequations}
\label{eq:mapping 4 ferm to 4 qubits}
\begin{align}
    \fket{\!0\!}    &\mapsto\,e^{i\phi_{0}}\,\ket{\!0000\!}\,,\\
    \fket{\!1_{1}\!}\!\fket{\!1_{2}\!}   &\mapsto\,e^{i\phi_{12}}\,\ket{\!1100\!}\,,\\
    \fket{\!1_{1}\!}\!\fket{\!1_{3}\!}   &\mapsto\,e^{i\phi_{13}}\,\ket{\!1010\!}\,,\\
    \fket{\!1_{1}\!}\!\fket{\!1_{4}\!}   &\mapsto\,e^{i\phi_{14}}\,\ket{\!1001\!}\,,\\
    \fket{\!1_{2}\!}\!\fket{\!1_{3}\!}   &\mapsto\,e^{i\phi_{23}}\,\ket{\!0110\!}\,,\\
    \fket{\!1_{2}\!}\!\fket{\!1_{4}\!}   &\mapsto\,e^{i\phi_{24}}\,\ket{\!0101\!}\,,\\
    \fket{\!1_{3}\!}\!\fket{\!1_{4}\!}   &\mapsto\,e^{i\phi_{34}}\,\ket{\!0011\!}\,,\\
    \fket{\!1_{1}\!}\!\fket{\!1_{2}\!}\fket{\!1_{3}\!}\!\fket{\!1_{4}\!}   &\mapsto\,e^{i\phi_{1234}}\,\ket{\!1111\!}\,.
\end{align}
\end{subequations}
\vspace*{1mm}
%\newpage
\newpage
\begin{table}[ht!]
\setlength{\tabcolsep}{7pt}
\renewcommand{\arraystretch}{1.5}
\begin{tabular}{c c c}
    \hline\hline
    $i,j$    &   $\fbra{\!0\!}\rho_{i,j}\fket{\!1_{i}\!}\!\fket{\!1_{j}\!}$  &   $\fbra{\!1_{i}\!}\rho_{i,j}\fket{\!1_{j}\!}$ \\   \hline
    $1,2$   & $\alpha_{0}\alpha_{12}^{*}+\alpha_{34}\alpha_{1234}^{*}$  &  $\alpha_{13}\alpha_{23}^{*}+\alpha_{14}\alpha_{24}^{*}$  \\  \hline
    $3,4$   & $\alpha_{0}\alpha_{34}^{*}+\alpha_{12}\alpha_{1234}^{*}$  &  $\alpha_{13}\alpha_{14}^{*}+\alpha_{23}\alpha_{24}^{*}$  \\  \hline
    $1,3$   & $\alpha_{0}\alpha_{13}^{*}-\alpha_{24}\alpha_{1234}^{*}$  &  $\alpha_{14}\alpha_{34}^{*}-\alpha_{12}\alpha_{23}^{*}$  \\  \hline
    $2,4$   & $\alpha_{0}\alpha_{24}^{*}-\alpha_{13}\alpha_{1234}^{*}$  &  $\alpha_{12}\alpha_{14}^{*}-\alpha_{23}\alpha_{34}^{*}$  \\  \hline
    $1,4$   & $\alpha_{0}\alpha_{14}^{*}+\alpha_{23}\alpha_{1234}^{*}$  &  $-\alpha_{12}\alpha_{24}^{*}-\alpha_{13}\alpha_{34}^{*}$  \\  \hline
    $2,3$   & $\alpha_{0}\alpha_{23}^{*}+\alpha_{14}\alpha_{1234}^{*}$  &  $\alpha_{12}\alpha_{13}^{*}+\alpha_{24}\alpha_{34}^{*}$  \\  \hline
    \hline
\end{tabular}
\caption{The off-diagonal matrix elements of the reductions $\rho_{i,j}=\tr_{\lnot i,j}\bigl(\fket{\!\psi_{{\rm even}}^{(4)}\!}\!\fbra{\!\psi_{{\rm even}}^{(4)}\!}\bigr)$ are shown.}
\label{table:example state off diags}
\end{table}

Performing the mapping of~(\ref{eq:mapping 4 ferm to 4 qubits}) for the state of Eq.~(\ref{eq:four mode even pure state}), \ie, $\fket{\!\psi_{{\rm even}}^{(4)}\!}\mapsto\ket{\!\psi_{{\rm even}}^{(4)}\!}$, and taking the partial traces for the qubits as usual, we obtain the off-diagonal elements that are to be compared with those in Table~\ref{table:example state off diags}. For instance, comparing $|\bra{\!0000\!}\tr_{3,4}\bigl(\ket{\!\psi_{{\rm even}}^{(4)}\!}\!\bra{\!\psi_{{\rm even}}^{(4)}\!}\bigr)\ket{\!1100\!}|$ with $|\fbra{\!0\!}\rho_{1,2}\fket{\!1_{1}\!}\!\fket{\!1_{2}\!}|$ we get the condition
\begin{align}
    |e^{i(\phi_{0}-\phi_{12})}\alpha_{0}\alpha_{12}^{*}+e^{i(\phi_{34}-\phi_{1234})}\alpha_{34}\alpha_{1234}^{*}| \nonumber  \\
    \ \ =\,|\alpha_{0}\alpha_{12}^{*}+\alpha_{34}\alpha_{1234}^{*}|\,.
    \label{eq:condition i derivation}
\end{align}
Since this must hold independently of the values of $\alpha_{0}$, $\alpha_{12}$, $\alpha_{34}$, and $\alpha_{1234}$, we arrive at
\begin{align}
    \phi_{12}+\phi_{34}-\phi_{0}-\phi_{1234}    &=\,2\,n_{1}\,\pi\,,
    \label{eq:condition i}
\end{align}
where $n_{1}\in\mathbb{Z}$. Similarly, the other off-diagonals from Table~\ref{table:example state off diags} provide the conditions
\begin{subequations}
\label{eq:conditions ii to vi}
\begin{align}
    \phi_{13}+\phi_{24}-\phi_{0}-\phi_{1234}    &=\,(2\,n_{2}+1)\,\pi\,,
    \label{eq:condition ii}\\[1mm]
    \phi_{14}+\phi_{23}-\phi_{0}-\phi_{1234}    &=\,2\,n_{3}\,\pi\,,
    \label{eq:condition iii}\\[1mm]
    \phi_{23}+\phi_{14}-\phi_{13}-\phi_{24}    &=\,2\,n_{4}\,\pi\,,
    \label{eq:condition iv}\\[1mm]
    \phi_{12}+\phi_{34}-\phi_{14}-\phi_{23}    &=\,(2\,n_{5}+1)\,\pi\,,
    \label{eq:condition v}\\[1mm]
    \phi_{13}+\phi_{24}-\phi_{12}-\phi_{34}    &=\,2\,n_{6}\,\pi\,,
    \label{eq:condition vi}
\end{align}
\end{subequations}
with $n_{2,3,4,5,6}\in\mathbb{Z}$. These conditions cannot all be met at the same time. This can be seen, for instance, by combining~(\ref{eq:condition i}) with~(\ref{eq:condition iii}), and comparing to~(\ref{eq:condition v}), which results in
\begin{align}
    2(n_{1}-n_{3})  &=\,(2\,n_{5}+1)\,,
    \label{eq:i-iii compared with v}
\end{align}
which cannot be satisfied, since the left-hand side is an even integer for all~$n_{1},n_{3}\in\mathbb{Z}$, while the right-hand side is an odd integer for all~$n_{5}\in\mathbb{Z}$. We hence conclude that the state of Eq.~(\ref{eq:four mode even pure state}) cannot be consistently mapped to~a four-qubit state, even though it satisfies the SSR, and despite the fact that for any of its bipartitions the respective marginals have the same spectra.

%%
%\newpage


\begin{thebibliography}{99}

%\vspace*{-3mm}

%%[1]
\bibitem{FriisDunjkoDuerBriegel2014}
N.~Friis, V.~Dunjko, W.~D{\"u}r, and H.~J.~Briegel,\
\emph{Implementing quantum control for unknown subroutines},\
%%DOI:\ 10.1103/PhysRevA.89.030303
\href{http://dx.doi.org/10.1103/PhysRevA.89.030303}{Phys.}\
\href{http://dx.doi.org/10.1103/PhysRevA.89.030303}{Rev.\ A\ \textbf{89}, 030303(R) (2014)}\
[\href{http://arxiv.org/abs/1401.8128}{arXiv:1401.8128}].
%%IOP Style
%%Friis~N, Dunjko~V, D{\"u}r~W and Briegel~H~J\ 2014\ \textit{Phys.\ Rev.}\ A\ \href{http://dx.doi.org/10.1103/PhysRevA.89.030303}{\textbf{89} 030303(R)}\
%%\textit{Preprint}\ arXiv:\href{http://arxiv.org/abs/1401.8128}{1401.8128} [quant-ph]

%%[2]
\bibitem{SchliemannLossMacDonald2001}
J.~Schliemann, D.~Loss and A.~H.~MacDonald,\
\emph{Double-Occupancy Errors, Adiabaticity, and Entanglement of Spin-Qubits in Quantum Dots},\
%%DOI:\ 10.1103/PhysRevB.63.085311
\href{http://dx.doi.org/10.1103/PhysRevB.63.085311}{Phys.\ Rev.\ B\ \textbf{63}, 085311}
\href{http://dx.doi.org/10.1103/PhysRevB.63.085311}{(2001)}\
[\href{http://arxiv.org/abs/cond-mat/0009083}{arXiv:cond-mat/0009083}].
%%IOP Style
%%Schliemann~J, Loss~D and MacDonald~A~H\ 2001\ \textit{Phys.\ Rev.}\ B\ \href{http://dx.doi.org/10.1103/PhysRevB.63.085311}{\textbf{63} 085311}\
%%\textit{Preprint}\ arXiv:cond-mat/\href{http://arxiv.org/abs/cond-mat/0009083}{0009083}

%%[3]
\bibitem{SchliemannCiracKusLewensteinLoss2001}
J.~Schliemann, J.~I.~Cirac, M.~Ku{\'s}, M.~Lewenstein and D.~Loss,\
\emph{Quantum Correlations in Two-Fermion Systems},\
%%DOI:\ 10.1103/PhysRevA.64.022303
\href{http://dx.doi.org/10.1103/PhysRevA.64.022303}{Phys.\ Rev.\ A\ \textbf{64}, 022303 (2001)}\
[\href{http://arxiv.org/abs/quant-ph/0012094}{arXiv:quant-}\href{http://arxiv.org/abs/quant-ph/0012094}{ph/0012094}].
%%IOP Style
%%Schliemann~J, Cirac~J~I, Ku{\'s}~M, Lewenstein~M and Loss~D\ 2001\ \textit{Phys.\ Rev.}\ A\
%%\href{http://dx.doi.org/10.1103/PhysRevA.64.022303}{\textbf{64} 022303}\
%%\textit{Preprint}\ arXiv:quant-ph/\href{http://arxiv.org/abs/quant-ph/0012094}{0012094}

%%[4]
\bibitem{LiZengLiuLong2001}
Y.~S.~Li, B.~Zeng, X.~S.~Liu and G.~L.~Long,\
\emph{Entanglement in a two-identical-particle system},\
%%DOI:\ 10.1103/PhysRevA.64.054302
\href{http://dx.doi.org/10.1103/PhysRevA.64.054302}{Phys.\ Rev.\ A\ \textbf{64},}
\href{http://dx.doi.org/10.1103/PhysRevA.64.054302}{054302 (2001)}\
[\href{http://arxiv.org/abs/quant-ph/0104101}{arXiv:quant-ph/0104101}].
%%IOP Style
%%Li~Y~S, Zeng~B, Liu~X~S and Long~G~L\ 2001\ \textit{Phys.\ Rev.}\ A\ \href{http://dx.doi.org/10.1103/PhysRevA.64.054302}{\textbf{64} 054302}\
%%\textit{Preprint}\ arXiv:quant-ph/\href{http://arxiv.org/abs/quant-ph/0104101}{0104101}

%%[5]
\bibitem{EckertSchliemannBrussLewenstein2002}
K.~Eckert, J.~Schliemann, D.~Bru{\ss} and M.~Lewenstein,\
\emph{Quantum Correlations in Systems of Indistinguishable Particles},\
%%DOI:\ 10.1006/aphy.2002.6268
\href{http://dx.doi.org/10.1006/aphy.2002.6268}{Annals\ Phys. \textbf{299}, 88 (2002)}\
[\href{http://arxiv.org/abs/quant-ph/0203060}{arXiv:quant-}\href{http://arxiv.org/abs/quant-ph/0203060}{ph/0203060}].
%%IOP Style
%%Eckert~K, Schliemann~J, Bru{\ss}~D and Lewenstein~M\ 2002\ \textit{Annals\ Phys.}\
%%\href{http://dx.doi.org/10.1006/aphy.2002.6268}{\textbf{299} 88\textendash127}\
%%\textit{Preprint}\ arXiv:quant-ph/\href{http://arxiv.org/abs/quant-ph/0203060}{0203060}

%%[6]
\bibitem{Shi2003}
Y.~Shi,\
\emph{Quantum entanglement of identical particles},\
%%DOI:\ 10.1103/PhysRevA.67.024301
\href{http://dx.doi.org/10.1103/PhysRevA.67.024301}{Phys.\ Rev.\ A\ \textbf{67}, 024301 (2003)}\
[\href{http://arxiv.org/abs/quant-ph/0205069}{arXiv:quant-}\href{http://arxiv.org/abs/quant-ph/0205069}{ph/0205069}].
%%IOP Style
%%Shi~Y\ 2003\ \textit{Phys.\ Rev.}\ A\ \href{http://dx.doi.org/10.1103/PhysRevA.67.024301}{\textbf{67} 024301}\
%%\textit{Preprint}\ arXiv:quant-ph/\href{http://arxiv.org/abs/quant-ph/0205069}{0205069}

%%[7]
\bibitem{BoteroReznik2004}
A.~Botero and B.~Reznik,\
\emph{BCS-like Modewise Entanglement of Fermion Gaussian States},\
%%DOI:\ 10.1016/j.physleta.2004.08.037
\href{http://dx.doi.org/10.1016/j.physleta.2004.08.037}{Phys.\ Lett.\ A\ \textbf{331}, 39}
\href{http://dx.doi.org/10.1016/j.physleta.2004.08.037}{(2004)}\
[\href{http://arxiv.org/abs/quant-ph/0404176v3}{arXiv:quant-ph/0404176}].
%%IOP Style
%%Botero~A and Reznik~B\ 2004\ \textit{Phys.\ Lett.}\ A\
%%\href{http://dx.doi.org/10.1016/j.physleta.2004.08.037}{\textbf{331} 39\textendash44}\
%%\textit{Preprint}\ arXiv:quant-ph/\href{http://arxiv.org/abs/quant-ph/0404176v3}{0404176}

%%[8]
\bibitem{CabanPodlaskiRembielinskiSmolinskiWalczak2005}
P.~Caban, K.~Podlaski, J.~Rembieli{\'n}ski, K.~A.~Smoli{\'n}ksi, and Z.~Walczak,\
\emph{Entanglement and tensor product decomposition for two fermions},\
%%DOI:\ 10.1088/0305-4470/38/6/L02
\href{http://dx.doi.org/10.1088/0305-4470/38/6/L02}{J.\ Phys.\ A:\ Math.\ Gen.}\
\href{http://dx.doi.org/10.1088/0305-4470/38/6/L02}{\textbf{38}, L79%–L86
(2005)}\ [\href{http://arxiv.org/abs/quant-ph/0405108}{arXiv:quant-ph/0405108}].
%%IOP Style
%%Caban~P, Podlaski~K, Rembieli{\'n}ski~J, Smoli{\'n}ksi~K~A and Walczak~Z\ 2005\ \textit{J.\ Phys.}\ A:\ \textit{Math.\ Gen.}\
%%\href{http://dx.doi.org/10.1088/0305-4470/38/6/L02}{\textbf{38} L79\textendash86}\
%%\textit{Preprint}\ arXiv:quant-ph/\href{http://arxiv.org/abs/quant-ph/0405108}{0405108}

%%[9]
\bibitem{BanulsCiracWolf2007}
M.-C.~Ba\~{n}uls, J.~I.~Cirac, and M.~M.~Wolf,\
\emph{Entanglement in fermionic systems},\
%%DOI:\ 10.1103/PhysRevA.76.022311
%% e-print \href{http://arxiv.org/abs/0705.1103}{arXiv:0705.1103} [quant-ph] (2007).
\href{http://dx.doi.org/10.1103/PhysRevA.76.022311}{Phys.\ Rev.\ A\ \textbf{76}, 022311 (2007)}\
[\href{http://arxiv.org/abs/0705.1103}{arXiv:0705.1103}].
%%IOP Style
%%Ba\~{n}uls~M~C, Cirac~J~I, and Wolf~M~M\ 2007\
%%\textit{Phys.\ Rev.}\ A\ \href{http://dx.doi.org/10.1103/PhysRevA.76.022311}{\textbf{76} 022311}\
%%\textit{Preprint}\ arXiv:\href{http://arxiv.org/abs/0705.1103}{0705.1103} [quant-ph]

%%[10]
\bibitem{BalachandranGovindarajanDeQueirozReyesLega2013}
A.~P.~Balachandran, T.~R.~Govindarajan, A.~R. de~Queiroz, and A.~F.~Reyes-Lega,\
\emph{Entanglement and Particle Identity: A Unifying Approach},\
%%DOI:\ 10.1103/PhysRevLett.110.080503
\href{http://dx.doi.org/10.1103/PhysRevLett.110.080503}{Phys.\ Rev.\ Lett.}\
\href{http://dx.doi.org/10.1103/PhysRevLett.110.080503}{\textbf{110}, 080503 (2013)}\
[\href{http://arxiv.org/abs/1303.0688}{arXiv:1303.0688}].
%%IOP Style
%%Balachandran~A~P, Govindarajan~T~R, de~Queiroz~A~R and Reyes-Lega~A~F\ 2013\ \textit{Phys.\ Rev.\ Lett.}\
%%\href{http://dx.doi.org/10.1103/PhysRevLett.110.080503}{\textbf{110} 080503}\
%%\textit{Preprint}\ arXiv:\href{http://arxiv.org/abs/1303.0688}{1303.0688} [hep-th]

%%[11]
\bibitem{Moriya2002}
H.~Moriya,\
\emph{Some aspects of quantum entanglement for CAR systems},\
%%DOI:\ 10.1023/A:1016158125660
\href{http://dx.doi.org/10.1023/A:1016158125660}{Lett.\ Math.\ Phys.\ \textbf{60}, 109 (2002)}\
[\href{http://arxiv.org/abs/math-ph/0110022}{arXiv:math-ph/0110022}].
%%IOP Style
%%Moriya~H\ 2002\ \textit{Lett.\ Math.\ Phys.}\ \href{http://dx.doi.org/10.1023/A:1016158125660}{\textbf{60} 109\textendash21}\
%%\textit{Preprint}\ arXiv:math-ph/\href{http://arxiv.org/abs/math-ph/0110022}{0110022}

%%[12]
\bibitem{ArakiMoriya2003}
H.~Araki and H.~Moriya,\
\emph{Joint Extension of States of Subsystems for a CAR System},\
%%DOI:\ 10.1007/s00220-003-0832-6
\href{http://dx.doi.org/10.1007/s00220-003-0832-6}{Commun.\ Math.\ Phys.}\
\href{http://dx.doi.org/10.1007/s00220-003-0832-6}{\textbf{237}, 105 (2003)}\
[\href{http://arxiv.org/abs/math-ph/0306044}{arXiv:math-ph/0306044}].
%%IOP Style
%%Araki~H and Moriya~H\ 2003\ \textit{Commun.\ Math.\ Phys.}\
%%\href{http://dx.doi.org/10.1007/s00220-003-0832-6}{\textbf{237} 105\textendash22}\
%%\textit{Preprint}\ arXiv:math-ph/\href{http://arxiv.org/abs/math-ph/0306044}{0306044}

%%[13]
\bibitem{Moriya2005}
H.~Moriya,\
\emph{Validity and failure of some entropy inequalities for CAR systems},\
%%DOI:\ 10.1063/1.1850995
\href{http://dx.doi.org/10.1063/1.1850995}{J.\ Math.\ Phys.\ \textbf{46}, 033508 (2005)}\
[\href{http://arxiv.org/abs/math-ph/0405042}{arXiv:math-ph/0405042}].
%%IOP Style
%%Moriya~H\ 2005\ \textit{J.\ Math.\ Phys.}\ \href{http://dx.doi.org/10.1063/1.1850995}{\textbf{46} 033508}\
%%\tetxit{Preprint}\ arXiv:math-ph/\href{http://arxiv.org/abs/math-ph/0405042}{0405042}

%%[14]
\bibitem{Harlow2016}
D.~Harlow,\
\emph{Jerusalem Lectures on Black Holes and Quantum Information},\
%%DOI:\ 10.1103/RevModPhys.88.015002
\href{http://dx.doi.org/10.1103/RevModPhys.88.015002}{Rev.\ Mod.\ Phys.\ \textbf{88}, 015002 (2016)}\
[\href{http://arxiv.org/abs/1409.1231}{arXiv:1409.1231}].
%%IOP Style
%%Harlow~D\ 2016\ \textit{Rev.\ Mod.\ Phys.}\ \href{http://dx.doi.org/10.1103/RevModPhys.88.015002}{\textbf{88} 015002}\
%%\textit{Preprint}\ arXiv:\href{http://arxiv.org/abs/1409.1231}{1409.1231} [hep-th]

%%[15]
\bibitem{FriisLeeBruschi2013}
N.~Friis, A.~R.~Lee, and D.~E.~Bruschi,\
\emph{Fermionic mode entanglement in quantum information},\
%%DOI:\ 10.1103/PhysRevA.87.022338
%%e-print \href{http://arxiv.org/abs/1211.7217}{arXiv:1211.7217} [quant-ph] (2012).
\href{http://dx.doi.org/10.1103/PhysRevA.87.022338}{Phys.\ Rev.\ A\ \textbf{87},}\
\href{http://dx.doi.org/10.1103/PhysRevA.87.022338}{022338 (2013)}\
[\href{http://arxiv.org/abs/1211.7217}{arXiv:1211.7217}].
%%IOP Style
%%Friis~N, Lee~A~R and Bruschi~D~E\ 2013\ \textit{Phys.\ Rev.}\ A\ \href{http://dx.doi.org/10.1103/PhysRevA.87.022338}{\textbf{87} 022338}\
%%\textit{Preprint}\ arXiv:\href{http://arxiv.org/abs/1211.7217}{1211.7217} [quant-ph]

%%[16]
\bibitem{WickWightmanWigner1952}
G.~C.~Wick, A.~S.~Wightman, and E.~P.~Wigner,\
\emph{The Intrinsic Parity of Elementary Particles},\
%%DOI:\ 10.1103/PhysRev.88.101
\href{http://dx.doi.org/10.1103/PhysRev.88.101}{Phys.\ Rev.\ \textbf{88},}\
\href{http://dx.doi.org/10.1103/PhysRev.88.101}{101 (1952)}.
%%IOP Style
%%Wick~G~C, Wightman~A~S and Wigner~E~P\ 1952\
%%\textit{Phys.\ Rev.}\ \href{http://dx.doi.org/10.1103/PhysRev.88.101}{\textbf{88} 101\textendash5}

%%[17]
\bibitem{ChristensonCroninFitchTurlay1964}
J.~H.~Christenson, J.~W.~Cronin, V.~L.~Fitch, and R.~Turlay,\
\emph{Evidence for the $2\pi$ Decay of the $K_{\rm 2}^{\ \rm 0}$ Meson},\
%%DOI:\ 10.1103/PhysRevLett.13.138
%%Keyword: CP violation
\href{http://dx.doi.org/10.1103/PhysRevLett.13.138}{Phys.\ Rev.\ Lett.\ \textbf{13}, 138 (1964)}.
%%IOP Style
%%Christenson~J~H, Cronin~J~W, Fitch~V~L and Turlay~R\ 1964\ \textit{Phys.\ Rev.\ Lett.}\
%%\href{http://dx.doi.org/10.1103/PhysRevLett.13.138}{\textbf{13} 138}

%%[18]
\bibitem{AlaviHaratiEtAlKTeVCollaboration1999}
A.~Alavi-Harati et al. (KTeV Collaboration),\
\emph{Observation of Direct CP Violation in $K_{S,L}\rightarrow\pi\pi$ Decays},\
%%DOI:\ 10.1103/PhysRevLett.83.22
\href{http://dx.doi.org/10.1103/PhysRevLett.83.22}{Phys.\ Rev.}\
\href{http://dx.doi.org/10.1103/PhysRevLett.83.22}{Lett.\ \textbf{83}, 22 (1999)}\
[\href{http://arxiv.org/abs/hep-ex/9905060}{arXiv:hep-ex/9905060}].
%%IOP Style
%%Alavi-Harati~A et al.(KTeV Collaboration)\ 1999\ \textit{Phys.\ Rev.\ Lett.}\ \href{http://dx.doi.org/10.1103/PhysRevLett.83.22}{\textbf{83} 22\textendash7}\
%%\textit{Preprint}\ arXiv:hep-ex/\href{http://arxiv.org/abs/hep-ex/9905060}{9905060}

%%[19]
\bibitem{Lueders1954}
G.~L{\"u}ders,\
\emph{On the Equivalence of Invariance under Time Reversal and under Particle-Antiparticle Conjugation for Relativistic Field Theories},\
%%URL:\ http://www.sdu.dk/media/bibpdf/Bind%2020-29/Bind/mfm-28-5.pdf
Dan.\ Mat.\ Fys.\ Medd.\ \textbf{28}, 1 (1954).
%%IOP Style
%%L{\"u}ders~G\ 1954\ \textit{Dan.\ Mat.\ Fys.\ Medd.}\ \textbf{28}, 1

%%[20]
\bibitem{PauliRosenfeldWeisskopf1955}
W.~Pauli, L.~Rosenfeld, and V.~Weisskopf (eds.),\
\emph{Niels Bohr and the Development of Physics}\ (Pergamon Press, London, 1955).

%%[21]
\bibitem{Lueders1957}
G.~L{\"u}ders,\
\emph{Proof of the TCP theorem},\
%%DOI:\ 10.1016/0003-4916(57)90032-5
\href{http://dx.doi.org/10.1016/0003-4916(57)90032-5}{Annals\ Phys.\ \textbf{2}, 1}\
\href{http://dx.doi.org/10.1016/0003-4916(57)90032-5}{(1957)}.
%%IOP Style
%%L{\"u}ders~G\ 1957\ \textit{Annals\ Phys.}\ \href{http://dx.doi.org/10.1016/0003-4916(57)90032-5}{\textbf{2} 1\textendash15}

%%[22]
\bibitem{HegerfeldtKrausWigner1968}
G.~C.~Hegerfeldt, K.~Kraus, and E.~P.~Wigner,\
\emph{Proof of the Fermion Superselection Rule without the Assumption of Time-Reversal Invariance},\
%%DOI:\ 10.1063/1.1664539
\href{http://dx.doi.org/10.1063/1.1664539}{J.\ Math.\ Phys.\ \textbf{9}, 2029}
\href{http://dx.doi.org/10.1063/1.1664539}{(1968)}.
%%IOP Style
%%Hegerfeldt~G~C, Kraus~K and Wigner~E~P\ 1968\ \textit{J.\ Math.\ Phys.}\ \href{http://dx.doi.org/10.1063/1.1664539}{\textbf{9} 2029}

%%[23]
\bibitem{ChiribellaDArianoPerinotti2010}
G.~Chiribella, G.~M.~D'Ariano, and P.~Perinotti,\
\emph{Probabilistic theories with purification},\
%%DOI:\ 10.1103/PhysRevA.81.062348
\href{http://dx.doi.org/10.1103/PhysRevA.81.062348}{Phys.\ Rev.\ A\ \textbf{81},}\
\href{http://dx.doi.org/10.1103/PhysRevA.81.062348}{062348 (2010)}\
[\href{http://arxiv.org/abs/0908.1583}{arXiv:0908.1583}].
%%IOP Style
%%Chiribella~G, D'Ariano~G~M and Perinotti~P\ 2010\ \textit{Phys.\ Rev.}\ A\
%%\href{http://dx.doi.org/10.1103/PhysRevA.81.062348}{\textbf{81} 062348}\
%%\textit{Preprint}\ arXiv:\href{http://arxiv.org/abs/0908.1583}{0908.1583} [quant-ph]

%%[24]
\bibitem{ChiribellaDArianoPerinotti2011}
G.~Chiribella, G.~M.~D'Ariano, and P.~Perinotti,\
\emph{Informational derivation of Quantum Theory},\
%%DOI:\ 10.1103/PhysRevA.84.012311
\href{http://dx.doi.org/10.1103/PhysRevA.84.012311}{Phys.\ Rev.\ A}\
\href{http://dx.doi.org/10.1103/PhysRevA.84.012311}{\textbf{84}, 012311 (2011)}\
[\href{http://arxiv.org/abs/1011.6451}{arXiv:1011.6451}].
%%IOP Style
%%Chiribella~G, D'Ariano~G~M and Perinotti~P\ 2011\ \textit{Phys.\ Rev.}\ A\
%%\href{http://dx.doi.org/10.1103/PhysRevA.84.012311}{\textbf{84} 012311}\
%%\textit{Preprint}\ arXiv:\href{http://arxiv.org/abs/1011.6451}{1011.6451} [quant-ph]

%%[25]
\bibitem{ChiribellaDArianoPerinotti2012}
G.~Chiribella, G.~M.~D'Ariano, and P.~Perinotti,\
\emph{Quantum Theory, namely the pure and reversible theory of information},\
%%DOI:\ 10.3390/e14101877
\href{http://dx.doi.org/10.3390/e14101877}{Entropy\ \textbf{14}, 1877 (2012)}\
[\href{http://arxiv.org/abs/1209.5533}{arXiv:1209.5533}].
%%IOP Style
%%Chiribella~G, D'Ariano~G~M and Perinotti~P\ 2012\ \textit{Entropy}\
%%\href{http://dx.doi.org/10.3390/e14101877}{\textbf{14} 1877\textendash93}\
%%\textit{Preprint}\ arXiv:\href{http://arxiv.org/abs/1209.5533}{1209.5533} [quant-ph]

%%[26]
\bibitem{AharonovSusskind1967}
Y.~Aharonov and L.~Susskind,\
\emph{Charge Superselection Rule},\
%%DOI:\ 10.1103/PhysRev.155.1428
\href{http://dx.doi.org/10.1103/PhysRev.155.1428}{Phys.\ Rev.\ \textbf{155}, 1428 (1967)}.
%%IOP Style
%%Aharonov~Y and Susskind~L\ 1967\ \textit{Phys.\ Rev.}\
%%\href{http://dx.doi.org/10.1103/PhysRev.155.1428}{\textbf{155} 1428\textendash31}

%%[27]
\bibitem{Yoshihuku1972}
Y.~Yoshihuku,\
\emph{Critical Comment on the Nonexistence of the Charge Superselection Rule},\
%%DOI:\ 10.1143/PTP.47.2085
\href{http://dx.doi.org/10.1143/PTP.47.2085}{Prog.\ Theor.\ Phys.\ \textbf{47},}\
\href{http://dx.doi.org/10.1143/PTP.47.2085}{2085 (1972)}.
%%IOP Style
%%Yoshihuku~Y\ 1972\ \textit{Prog.\ Theor.\ Phys.}\
%%\href{http://dx.doi.org/10.1143/PTP.47.2085}{\textbf{47} 2085\textendash9}

%%[28]
\bibitem{StrocchiWightman1974}
F.~Strocchi and A.~S.~Wightman,\
\emph{Proof of the charge superselection rule in local relativistic quantum field theory},\
%%DOI:\ 10.1063/1.1666601
%%e-print \href{}{} [] ().
\href{http://dx.doi.org/10.1063/1.1666601}{J.\ Math.\ Phys.\ \textbf{15}, 2198 (1974)}.

%%[29]
\bibitem{VerstraeteCirac2003}
F.~Verstraete and J.~I.~Cirac,\
\emph{Quantum nonlocality in the presence of superselection rules and data hiding protocols},\
%%DOI:\ 10.1103/PhysRevLett.91.010404
\href{http://dx.doi.org/10.1103/PhysRevLett.91.010404}{Phys.\ Rev.\ Lett.\ \textbf{91}, 010404 (2003)}\
[\href{http://arxiv.org/abs/quant-ph/0302039}{arXiv:quant-}
\href{http://arxiv.org/abs/quant-ph/0302039}{ph/0302039}].
%%IOP Style
%%Verstraete~F and Cirac~J~I\ 2003\ \textit{Phys.\ Rev.\ Lett.}\ \href{http://dx.doi.org/10.1103/PhysRevLett.91.010404}{\textbf{91} 010404}\
%%\textit{Preprint}\ arXiv:quant-ph/\href{http://arxiv.org/abs/quant-ph/0302039}{0302039}

%%[30]
\bibitem{BartlettWiseman2003}
S.~D.~Bartlett and H.~M.~Wiseman,\
\emph{Entanglement Constrained by Superselection Rules},\
%%DOI:\ 10.1103/PhysRevLett.91.097903
\href{http://dx.doi.org/10.1103/PhysRevLett.91.097903}{Phys.\ Rev.\ Lett.\ \textbf{91},}
\href{http://dx.doi.org/10.1103/PhysRevLett.91.097903}{097903 (2003)}\
[\href{http://arxiv.org/abs/quant-ph/0303140}{arXiv:quant-ph/0303140}].
%%IOP Style
%%Bartlett~S~D and Wiseman~H~M\ 2003\ \textit{Phys.\ Rev.\ Lett.}\ \href{http://dx.doi.org/10.1103/PhysRevLett.91.097903}{\textbf{91} 097903}\
%%\textit{Preprint}\ arXiv:quant-ph/\href{http://arxiv.org/abs/quant-ph/0303140}{0303140}

%%[31]
\bibitem{SchuchVerstraeteCirac2004}
N.~Schuch, F.~Verstraete, and J.~I.~Cirac,\
\emph{Nonlocal Resources in the Presence of Superselection Rules},\
%%DOI:\ 10.1103/PhysRevLett.92.087904
\href{http://dx.doi.org/10.1103/PhysRevLett.92.087904}{Phys.}\
\href{http://dx.doi.org/10.1103/PhysRevLett.92.087904}{Rev.\ Lett.\ \textbf{92}, 087904 (2004)}\
[\href{http://arxiv.org/abs/quant-ph/0310124}{arXiv:quant-ph/0310124}].
%%IOP Style
%%Schuch~N, Verstraete~F, and Cirac~J~I\ 2004\ \textit{Phys.\ Rev.\ Lett.}\
%%\href{http://dx.doi.org/10.1103/PhysRevLett.92.087904}{\textbf{92} 087904}\
%%\textit{Preprint}\ arXiv:quant-ph/\href{http://arxiv.org/abs/quant-ph/0310124}{0310124}

%%[32]
\bibitem{SkotioniotisTolouiDurhamSanders2013}
M.~Skotiniotis, B.~Toloui, I.~T.~Durham, and B.~C.~Sanders,\
\emph{Quantum frameness for Charge-Parity-Time inversion symmetry},\
%%DOI:\ 10.1103/PhysRevLett.111.020504
\href{http://dx.doi.org/10.1103/PhysRevLett.111.020504}{Phys.\ Rev.\ Lett.\ \textbf{111}, 020504}
\href{http://dx.doi.org/10.1103/PhysRevLett.111.020504}{(2013)}\
[\href{http://arxiv.org/abs/1306.6114}{arXiv:1306.6114}].
%%IOP Style
%%Skotiniotis~M, Toloui~B, Durham~I~T and Sanders~B~C\ 2013\ \textit{Phys.\ Rev.\ Lett.}\
%%\href{http://dx.doi.org/10.1103/PhysRevLett.111.020504}{\textbf{111} 020504}\
%%\textit{Preprint}\ arXiv:\href{http://arxiv.org/abs/1306.6114}{1306.6114} [quant-ph]

%%[33]
\bibitem{SkotioniotisTolouiDurhamSanders2014}
M.~Skotiniotis, B.~Toloui, I.~T.~Durham, and B.~C.~Sanders,\
\emph{Quantum Resource Theory for Charge-Parity-Time Inversion},\
%%DOI:\ 10.1103/PhysRevA.90.012326
\href{http://dx.doi.org/10.1103/PhysRevA.90.012326}{Phys.\ Rev.\ A\ \textbf{90}, 012326 (2014)}\
[\href{http://arxiv.org/abs/1405.2516}{arXiv:1405.2516}].
%%IOP Style
%%Skotiniotis~M, Toloui~B, Durham~I~T and Sanders~B~C\ 2014\ \textit{Phys.\ Rev.}\ A\
%%\href{http://dx.doi.org/10.1103/PhysRevA.90.012326}{\textbf{90} 012326}\
%%\textit{Preprint}\ arXiv:\href{http://arxiv.org/abs/1405.2516}{1405.2516} [quant-ph]

%%[34]
\bibitem{Friis:PhD2013}
N.~Friis,\
\emph{Cavity mode entanglement in relativistic quantum information},\
\href{http://etheses.nottingham.ac.uk/3795/}{Ph.D. thesis, University of Nottingham,}\
\href{http://etheses.nottingham.ac.uk/3795/}{2013}\
[\href{http://arxiv.org/abs/1311.3536}{arXiv:1311.3536}].

%%[35]
\bibitem{WisemanVaccaro2003}
H.~M.~Wiseman and J.~A.~Vaccaro,\
\emph{Entanglement of Indistinguishable Particles Shared between Two Parties},\
%%DOI:\ 10.1103/PhysRevLett.91.097902
\href{http://dx.doi.org/10.1103/PhysRevLett.91.097902}{Phys.\ Rev.\ Lett.\ \textbf{91}, 097902 (2003)}\
[\href{http://arxiv.org/abs/quant-ph/0210002}{arXiv:quant-}
\href{http://arxiv.org/abs/quant-ph/0210002}{ph/0210002}].
%%IOP Style
%%Wiseman~H~M and Vaccaro~J~A\ 2003\ \textit{Phys.\ Rev.\ Lett.}\ \href{http://dx.doi.org/10.1103/PhysRevLett.91.097902}{\textbf{91} 097902}\
%%\textit{Preprint}\ arXiv:quant-ph/\href{http://arxiv.org/abs/quant-ph/0210002}{0210002}

%%[36]
\bibitem{WisemanBartlettVaccaro2004}
H.~M.~Wiseman, S.~D.~Bartlett, and J.~A.~Vaccaro,\
\emph{Ferreting out the Fluffy Bunnies: Entanglement constrained by Generalized superselection rules},\
%%DOI:\ 10.1142/9789812703002_0047
%%e-print \href{http://arxiv.org/abs/quant-ph/0309046}{arXiv:quant-ph/0309046} (2003).
\href{http://dx.doi.org/10.1142/9789812703002_0047}{Laser Spec-}
\href{http://dx.doi.org/10.1142/9789812703002_0047}{troscopy, Proceedings of the XVI International Con-}
\href{http://dx.doi.org/10.1142/9789812703002_0047}{ference, pp.~307 %-314
(World Scientific, 2004)}\ [\href{http://arxiv.org/abs/quant-ph/0309046}{arXiv:quant-}
\href{http://arxiv.org/abs/quant-ph/0309046}{ph/0309046}].
%%IOP Style
%%Wiseman~H~M, Bartlett~S~D and Vaccaro~J~A\ 2004\ in\ \textit{Laser Spectroscopy, Proceedings of the XVI International Conference}\ %%\href{http://dx.doi.org/10.1142/9789812703002_0047}{307\textendash14}
%%(World Scientific)\ \textit{Preprint}\ arXiv:quant-ph/\href{http://arxiv.org/abs/quant-ph/0309046}{0309046}

%%[37]
\bibitem{Pauli1940}
W.~Pauli,\
\emph{The Connection Between Spin and Statistics},\
%%DOI:\ 10.1103/PhysRev.58.716
\href{http://dx.doi.org/10.1103/PhysRev.58.716}{Phys.\ Rev.\ \textbf{58}, 716 (1940)}.
%%IOP Style
%%Pauli~W\ 1940\ \textit{Phys.\ Rev.}\ \href{http://dx.doi.org/10.1103/PhysRev.58.716}{\textbf{58} 716}

%%[38]
\bibitem{Schwinger1951}
J.~Schwinger,\
\emph{The Theory of Quantized Fields. I},\
%%DOI:\ 10.1103/PhysRev.82.914
%%Keywords: spin-statistics, CPT theorem
\href{http://dx.doi.org/10.1103/PhysRev.82.914}{Phys.}\
\href{http://dx.doi.org/10.1103/PhysRev.82.914}{Rev.\ \textbf{82}, 914 (1951)}.
%%IOP Style
%%Schwinger~J\ 1951\ \textit{Phys.\ Rev.}\ \href{http://dx.doi.org/10.1103/PhysRev.82.914}{\textbf{82} 914\textendash27}

%%[39]
\bibitem{PeskinSchroeder1995}
M.~E.~Peskin and D.~V.~Schroeder,\
\emph{An Introduction to Quantum Field Theory}\
(Westview Press, 1995).

%%[40]
\bibitem{EislerZimboras2015}
V.~Eisler and Z.~Zimbor{\'a}s,\
\emph{On the partial transpose of fermionic Gaussian states},\
%%DOI:\ 10.1088/1367-2630/17/5/053048
\href{http://dx.doi.org/10.1088/1367-2630/17/5/053048}{New\ J.\ Phys.\ \textbf{17}, 053048}
\href{http://dx.doi.org/10.1088/1367-2630/17/5/053048}{(2015)}\
[\href{http://arxiv.org/abs/1502.01369}{arXiv:1502.01369}].
%%IOP Style
%%Eisler~V and Zimbor{\'a}s~Z\ 2015\ \textit{New\ J.\ Phys.}\ \href{http://dx.doi.org/10.1088/1367-2630/17/5/053048}{\textbf{17} 053048}\
%%\textit{Preprint}\ arXiv:\href{http://arxiv.org/abs/1502.01369}{1502.01369} [cond-mat.stat-mech]

%%[41]
\bibitem{Barends-Martinis2015}
R.~Barends, L.~Lamata, J.~Kelly, L.~Garc{\'i}a-{\'A}lvarez, A.~G.~Fowler, A.~Megrant, E.~Jeffrey, T.~C.~White, D.~Sank, J.~Y.~Mutus, B.~Campbell, Yu~Chen, Z.~Chen, B.~Chiaro, A.~Dunsworth, I.-C.~Hoi, C.~Neill, P.~J.~J.~O'Malley, C.~Quintana, P.~Roushan, A.~Vainsencher, J.~Wenner, E.~Solano, and J.~M.~Martinis,\
\emph{Digital quantum simulation of fermionic models with a superconducting circuit},\
\href{http://dx.doi.org/10.1038/ncomms8654}{Nat.\ Commun.\ \textbf{6}, 7654 (2015)}\
[\href{http://arxiv.org/abs/1501.07703}{arXiv:1501.07703}].
%%IOP Style
%%Barends~R, Lamata~L, Kelly~J, Garc{\'i}a-{\'A}lvarez~L, Fowler~A~G, Megrant~A, Jeffrey~E, White~T~C,
%%Sank~D, Mutus~J~Y, Campbell~B, Chen~Y, Chen~Z, Chiaro~B, Dunsworth~A, Hoi~I-C, Neill~C, O'Malley~P~J~J,
%%Quintana~C, Roushan~P, Vainsencher~A, Wenner~J, Solano~E and Martinis~J~M\ 2015\
%%\textit{Nat.\ Commun.}\ \href{http://dx.doi.org/10.1038/ncomms8654}{\textbf{6} 7654}\
%%\textit{Preprint}\ arXiv:\href{http://arxiv.org/abs/1501.07703}{1501.07703} [quant-ph]

%%[42]
\bibitem{IorioLambiase2013}
A.~Iorio and G.~Lambiase,\
\emph{Quantum field theory in curved graphene spacetimes, Lobachevsky geometry, Weyl symmetry, Hawking effect, and all that},\
%%DOI:\ 10.1103/PhysRevD.90.025006
%%e-print \href{http://arxiv.org/abs/1308.0265}{arXiv:1308.0265} [hep-th] (2013),
\href{http://dx.doi.org/10.1103/PhysRevD.90.025006}{Phys.\ Rev.\ D}\
\href{http://dx.doi.org/10.1103/PhysRevD.90.025006}{\textbf{90}, 025006 (2014)}\
[\href{http://arxiv.org/abs/1308.0265}{arXiv:1308.0265}].
%%IOP Style
%%Iorio~A and Lambiase~G\ 2014\ \textit{Phys.\ Rev.}\ D\ \href{http://dx.doi.org/10.1103/PhysRevD.90.025006}{\textbf{90} 025006}\
%%\textit{Preprint}\ arXiv:\href{http://arxiv.org/abs/1308.0265}{1308.0265} [hep-th]

%%[43]
\bibitem{Iorio2015}
A.~Iorio,\
\emph{Curved Spacetimes and Curved Graphene: a status report of the Weyl-symmetry approach},\
%%DOI:\ 10.1142/S021827181530013X
%%e-print \href{http://arxiv.org/abs/1412.4554}{arXiv:1412.4554} [hep-th] (2014).
\href{http://dx.doi.org/10.1142/S021827181530013X}{Int.\ J.\ Mod.}\
\href{http://dx.doi.org/10.1142/S021827181530013X}{Phys.\ D\ \textbf{4}, 1530013 (2015)}\
[\href{http://arxiv.org/abs/1412.4554}{arXiv:1412.4554}].
%%IOP Style
%%Iorio~A\ 2015\ \textit{Int.\ J.\ Mod.\ Phys.}\ D\ \href{http://dx.doi.org/10.1142/S021827181530013X}{\textbf{4} 1530013}\
%%\textit{Preprint}\ arXiv:\href{http://arxiv.org/abs/1412.4554}{1412.4554} [hep-th]


\end{thebibliography}
\end{document}